\pdfoutput=1 

\documentclass{article}
\usepackage[T1]{fontenc}
\usepackage[utf8]{inputenc}
\usepackage[style=chem-angew, backend=biber, natbib=true, sorting=none]{biblatex} %
\usepackage[margin=1in]{geometry}
\usepackage[hidelinks]{hyperref}
\usepackage{color}
\usepackage{graphicx}
\usepackage[table,dvipsnames]{xcolor}
\usepackage{enumerate}
\usepackage{setspace}
\usepackage{mhchem}
\usepackage{url}
\usepackage{framed}
\usepackage{todonotes}
\usepackage{placeins}
\usepackage{longtable}
\usepackage{graphbox}
\usepackage{caption}
\usepackage{soul} %
\usepackage{tikz-cd}

\tikzcdset{column sep/normal=1.2em} %

\author{Connor W. Coley\thanks{Department of Chemical Engineering, Massachusetts Institute of Technology, Cambridge, MA 02139}~\footnote{ccoley@mit.edu}, ~ Natalie S. Eyke\footnotemark[1], ~ Klavs F. Jensen\footnotemark[1]~\footnote{kfjensen@mit.edu}}

\date{~}

\newcolumntype{L}[1]{>{\raggedright\let\newline\\\arraybackslash\hspace{0pt}}m{#1}}
\newcolumntype{C}[1]{>{\centering\let\newline\\\arraybackslash\hspace{0pt}}m{#1}}
\newcolumntype{R}[1]{>{\raggedleft\let\newline\\\arraybackslash\hspace{0pt}}m{#1}}

\newcommand{\onecolumnsize}{8.4cm}
\newcommand{\twocolumnsize}{17.8cm}

\newcommand{\lnameref}[1]{%
\bgroup
\let\nmu\MakeLowercase
\nameref{#1}\egroup}

\DeclareCiteCommand{\citenum}
  {}
  {\printfield{labelnumber}}
  {}
  {}

\title{Autonomous discovery in the chemical sciences part I: Progress}

\newif\iffull
\fulltrue

\begin{document}
\maketitle


\clearpage

\tableofcontents
\doublespacing
\newpage

\section{Abstract} 

This two-part review examines how automation has contributed to different aspects of discovery in the chemical sciences. In this first part, we describe a classification for discoveries of physical matter (molecules, materials, devices), processes, and models and how they are unified as search problems. We then introduce a set of questions and considerations relevant to assessing the extent of autonomy. Finally, we describe many case studies of discoveries accelerated by or resulting from computer assistance and automation from the domains of synthetic chemistry, drug discovery, inorganic chemistry, and materials science. These illustrate how rapid advancements in hardware automation and machine learning continue to transform the nature of experimentation and modelling.

Part two reflects on these case studies and identifies a set of open challenges for the field.

\section{Introduction}
The prospect of a robotic scientist has long been an object of curiosity, optimism, skepticism, and job-loss fear, depending on who is asked. As computing was becoming mainstream, excitement grew around the potential for logic and reasoning--the underpinnings of the scientific process--to be codified into computer programs; as hardware automation became more robust and cost effective, excitement grew around the potential for a universal synthesis platform to enhance the work of human chemists in the lab; and as data availability and statistical analysis/inference techniques improved, excitement grew around the potential for statistical models (machine learning included) to draw new insights from vast quantities of chemical information \cite{turing_i.computing_1950,bradshaw_studying_1983,turing_computers_1995,langley_computational_2000,valdes-perez_principles_1999,sparkes_towards_2010,sozou_computational_2017}. 

The confluence of these factors makes that prospect increasingly realistic. In organic chemistry, we have already seen proof-of-concept examples of the ``robo-chemist'' \cite{peplow_organic_2014} able to intelligently select and conduct experiments \cite{houben_automatic_2015, fitzpatrick_novel_2016, reizman_feedback_2016}; there have even been strides made toward a universal synthesis platform \cite{godfrey_remote-controlled_2013,li_synthesis_2015,lowe_derek_automated_2018,steiner_organic_2018}, theoretically capable of executing most chemical processes but highly constrained in practice. While there have been fewer success stories in automating drug discovery holistically \cite{schneider_automating_2018}, the excitement around machine learning in this application space is especially apparent, with dozens of start-up companies promising to revolutionize the development of new medicines through artificial intelligence \cite{smith_141_nodate}. 

A more pessimistic view of automated discovery is that machines will never be able to make real ``revolutions'' in science because they necessarily operate within a specific set of instructions  \cite{anderson_machines_2009}. This attitude is exemplified by {\it Lady Lovelace's objection}: ``The Analytical Engine has no pretensions to originate anything. It can do whatever we know how to order it to perform. It can follow analysis; but it has no power of anticipating any analytical relations or truths'' \cite{turing_i.computing_1950}. Some have expressed a milder sentiment, perhaps in light of advances in computing, cautioning that an increasing reliance on robotic tools might reduce the odds of a serendipitous discovery \cite{sparkes_towards_2010}. Muggleton is more declarative, stating that ``science is an essentially human activity that requires clarity both in the statement of hypotheses and their clear and undeniable refutation through experimentation'' \cite{muggleton_2020_2006}. However, there is little disagreement that automation and computation in science has improved productivity through efficiency, reduction of error, and the ability to address large-scale problems \cite{waltz_automating_2009}.

In the remainder of Part 1, we will discuss the different types of discovery typically reported in the chemical sciences and how they can be unified as searches in a high-dimensional design space. %
Along with this definition comes a recommended set of questions to ask when evaluating the extent to which a discovery can be attributed to automation or autonomy. 
We will then discuss a number of case studies arranged in terms of the type of discovery being pursued and the nature of the approach used to do so. 
Part 2 will reflect on these case studies and make explicit what we believe to be the primary obstacles to autonomous discovery.

\section{Defining discovery}
\subsection{Classifications of discoveries}

There is no single definition of what constitutes a scientific discovery.  \citeauthor{valdes-perez_principles_1999} defines discovery as ``the generation of novel, interesting, plausible, and intelligible knowledge'' \cite{valdes-perez_principles_1999}. Data-driven knowledge discovery, specifically, has been defined as the ``nontrivial extraction of implicit, previously unknown, and potential useful information'' \cite{Piateski:1991:KDD:583310}. Each of these criteria, however, is inherently subjective. 
``Novel'' is simultaneously ambiguous and considered distinct from ``new''; it is generally meant to indicate some level of nonobviousness or, by one definition, a lack of predictability \cite{Gromski2019}. However, if we artificially limit what we consider to be known and demonstrate a successful extrapolation to a conclusion that really \emph{was} known, it would be reasonable to argue that this does not constitute a discovery. This connects to the question of what it might mean for a discovery to be ``interesting'' or ``useful'', for which we avoid providing a precise definition.

\begin{figure}[h]
  \centering
  \includegraphics[width=10cm]{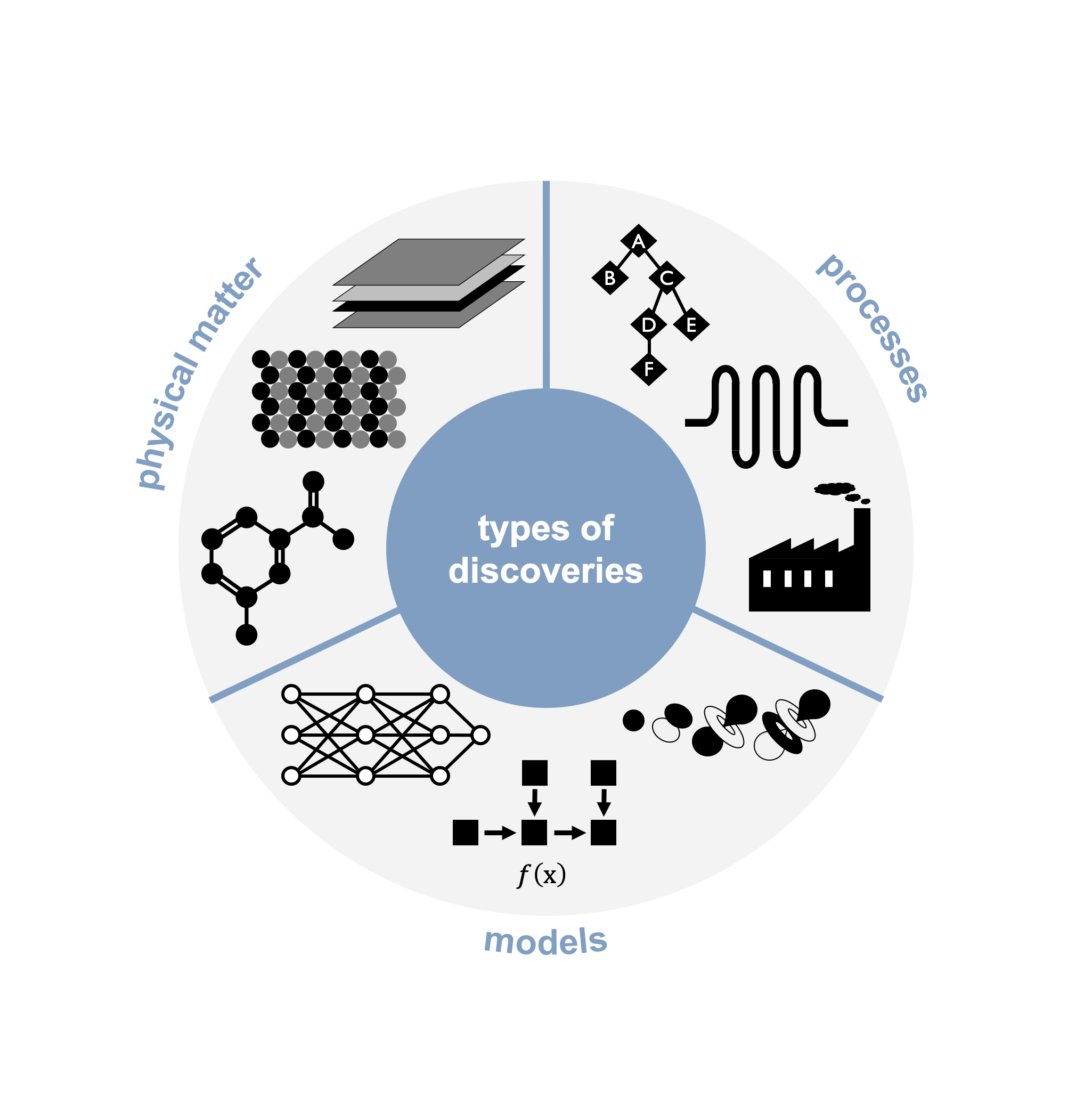}
  \caption{The three broad categories of discovery described in this review: physical matter, processes, and models.}
  \label{fig:our_types_discovery}
  \end{figure}

For the purposes of this review, we instead define three broad types of discoveries in the chemical sciences (Figure \ref{fig:our_types_discovery}) and provide examples of each.

\textbf{Physical matter.} Often, the ultimate result of a discovery campaign is the identification of a molecule (not discounting macromolecules), material, or device that achieves a desired function. This category encompasses most drug discovery efforts, where the output may be new chemical matter that could later become part of a therapeutic, as well as materials discovery for a wide array of applications.%

\textbf{Processes.} Discoveries may also take the form of processes. These may be abstract, like  the Haber-Bosch process, pasteurization, and directed evolution. They are more often concrete, like synthetic routes to organic molecules or a specific set of reaction conditions to achieve a chemical transformation. %

\textbf{Models.} Our definition of a model includes empirical models (such as those obtained through regression of experimental data), structure-function relationships, symbolic regressions, natural laws, and even conceptual models that provide mechanistic understanding. It is common for models to be \emph{part} of the discovery of the other two types as surrogates for experiments, as will be seen in many examples below.

The most famous examples of scientific discoveries in chemistry tend to be natural laws or theories that are able to rationalize observed phenomena that previous theories could not. Mendeleev's periodic table of the elements, Thomson's discovery of the electron, Rutherford's discovery of atomic nuclei, the Schrodinger equation, Kekul\'e's structure of benzene, et cetera. In their time, these represented radical departures from previous frameworks. Though we do consider these to be models, identifying them through computational or algorithmic approaches would require substantially more open-ended hypothesis generation than what is currently possible.  %

\subsection{Discovery as a search}

We argue that the process of scientific discovery can always be thought of as a search problem, regardless of the nature of that discovery \cite{langley_scientific_1987,klahr_dual_1988, sozou_computational_2017}.

Molecular discovery is a search within ``chemical space'' \cite{oprea_chemography:_2001,lipinski_navigating_2004,dobson_chemical_2004,reymond_chemical_2015}--an enormous combinatorial design space of theoretically-possible molecules. A common estimate of its size, considering only molecules of a limited size made up of CHONS atoms, is $10^{60}$ \cite{bohacek_art_1996}, although for any one application or with reasonable restrictions (e.g., on drug-likeness or synthetic accessibility), the size of the \emph{relevant} chemical space will be significantly smaller \cite{drew_size_2012,Walters2018}. Biological compounds exist in an even larger space if one considers that there are, e.g., $20^{100}$ theoretically-possible 100-peptide proteins using only canonical amino acids, although again the number that are foldable and biologically relevant will be significantly smaller. Materials discovery is another combinatorial design space, where structural composition must be defined by both discrete variables (e.g., elemental identities) and continuous variables (e.g., stoichometric ratios) and processing conditions. The design space for a device is even larger, as it compounds the complexity of its constituent components with additional considerations about its geometry.

Discovering a chemical or physical process is the result of searching a design space defined by process variables and/or sequences of operations. For example, optimizing a chemical reaction for its yield might involve changing species' concentrations, the reaction temperature, and the residence time \cite{mcmullen_automated_2010}. It may also include selecting the identity of a catalyst as a discrete variable \cite{j.reizman_suzukimiyaura_2016}, or changing the order of addition \cite{denmark_catalytic_1995}. A new research workflow can be thought of as the identification of actions to be taken and their timing, such as the development of split-and-pool combinatorial chemistry for diversity-oriented synthesis \cite{schreiber_target-oriented_2000} or a screening and selection strategy for directed evolution \cite{arnold1998design}.

The majority of models that are ``discovered'', under our broad definition, are empirical relationships that come from data fitting. In these cases, the search space is well-defined once an input representation (e.g., a set of descriptors or parameters) and a model family (e.g., a linear model, a deep neural network) are selected. While this can present a massive search space  when considering all possible values of all learned parameters (e.g., for deep learning regression techniques), the final model is often the result of a simplified, \emph{local} search from a random initialization (e.g., using stochastic gradient descent). Symbolic regressions are searches in a combinatorial space of input variables and mathematical operations \cite{schmidt_distilling_2009}. More abstract models, like mechanistic explanations of natural phenomena, exist in a high-dimensional hypothesis space that is difficult to formalize; automated discovery tools that are able to generate causal explanations do so using simplified terminology and well-defined ontologies \cite{gyori_word_2017}.

In virtually every case of computer-assisted discovery, the actual search space is significantly larger than what the program or platform is allowed to explore. We might decide to focus our attention on a specific set of compounds (e.g., a fixed scaffold), a specific class of materials (e.g., perovskites), a specific step in a catalyst synthesis process with a finite number of tunable process variables (e.g., the temperature and time of an annealing step), or a specific hypothesis structure (e.g., categorizing a ligand's effect on a protein as an agonist, antagonist, promoter, etc.). Constraining the search space is one way of integrating domain expertise/intuition into the discovery process. Moreover, it can greatly simplify the search process and mitigate the practical challenges of automated validation and feedback.

\subsection{The role of validation and feedback}
The way that we navigate the search space in a discovery effort is often iterative. Classically, the discovery of physical matter, such as in lead optimization for drug discovery, is divided into stages of design, make, test. An analogous cycle for searching hypothesis space could be described as hypothesize, validate, revise beliefs.

\begin{figure}[h]
  \centering
  \includegraphics[width=\linewidth]{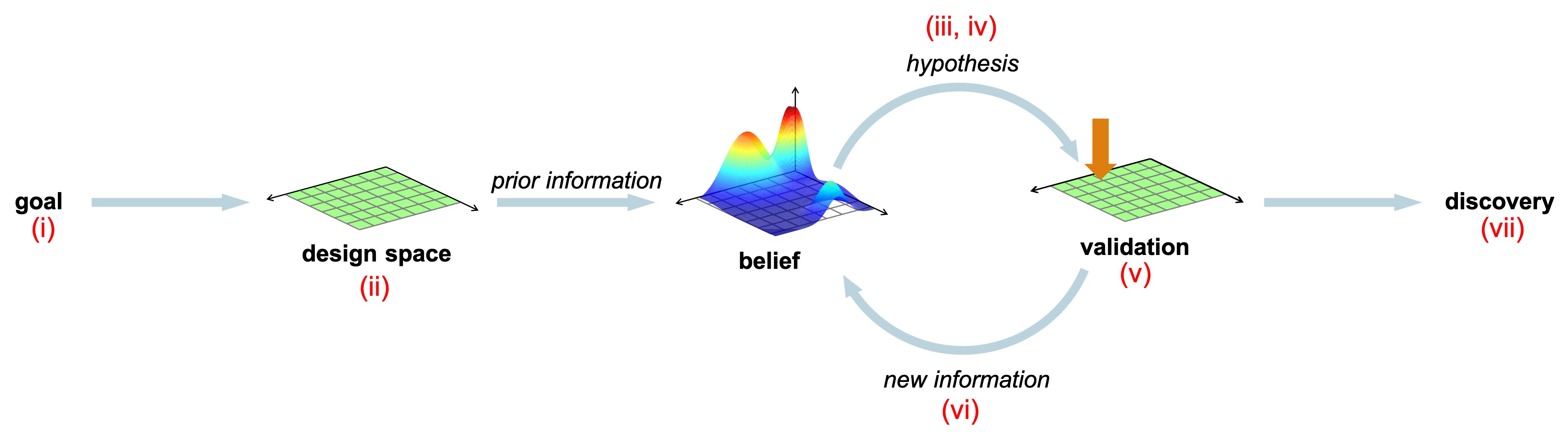}
  \caption{Simplified schematic of a hypothesis-driven (or model-driven) discovery process. When not proceeding iteratively, \emph{new information} is not used to revise our belief (current knowledge). Lowercase roman numerals ({\color{red} red}) correspond to the questions for assessing autonomy in discovery.}
 \label{fig:our_iterative_process_search}
  \end{figure}

This third step--test or revise beliefs--helps to explain the role of validation and feedback in discovery: experiments, physical or computational, serve to support or refute hypotheses. When information is imperfect or insufficient to lead to a confident prediction, it is important to collect new information  to improve our understanding of the problem. This might mean taking an empirical regression fit to a small number of data points, evaluating our uncertainty, and performing follow-up experiments to reduce our uncertainty in regions where we would like to have a more confident prediction (Figure~\ref{fig:our_iterative_process_search}). Purely virtual screening is not sufficient for drug discovery \cite{schneider_virtual_2010}, where experimental validation continues to be essential \cite{saikin_closed-loop_2019}; \citeauthor{Schneider2019} describe experimental testing of drugs designed using \emph{de novo} workflows as a ``non-negotiable'' criterion \cite{Schneider2019}. In the materials space as well,  \citeauthor{Halls2010} propose a materials discovery scheme in which synthesis, characterization, and testing are critical components \cite{Halls2010}. %
The scope of hypotheses that lend themselves to automated validation has limited the scope of discovery tasks that are able to be automated.

Consider a scenario where we have a large data set of molecular structures and a property of interest, like their \emph{in vitro} binding affinity for a particular protein target. We can perform a statistical regression to correlate the two and represent our understanding of the structure-function landscape. Based on that model, we may propose a new structure--a compound not yet tested--that is predicted to have high activity. Whether that constitutes discovery of the compound is ambiguous. Using scientific publication as the bar, it is reasonable to expect a high degree of confidence, regardless of whether that confidence arises from a statistical analysis of existing data or from confirmation through acquisition of new data. Even with a highly accurate model, performing a large virtual screen could lead to thousands of false positive results \cite{Walters2018}. For a philosophical discussion about the nature of knowledge and need for confidence, correctness, and justification, see the description of the Gettier problem in ref.~\citenum{gettier_is_1963}.

\begin{figure}[h]
  \centering
  \includegraphics[width=10cm]{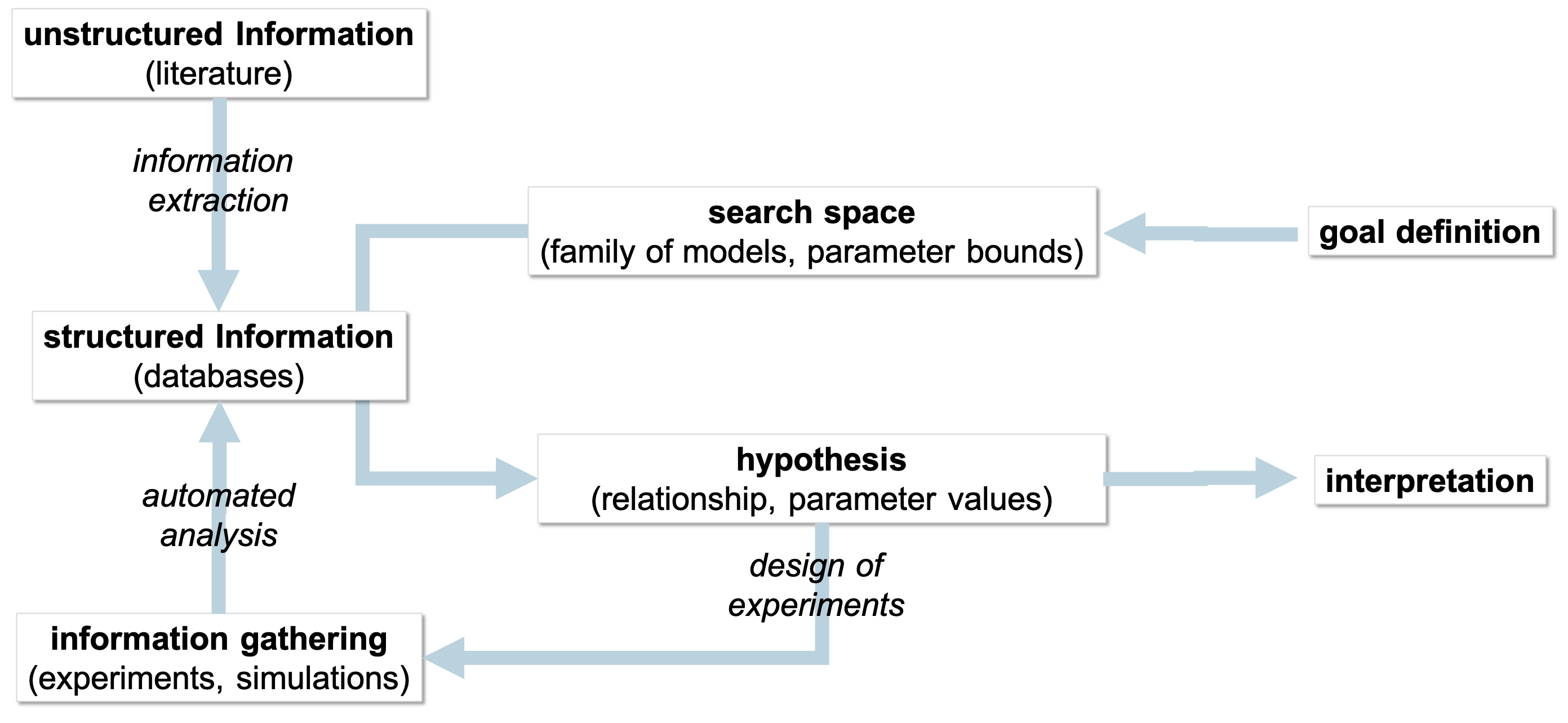}
  \caption{One way to visualize the discovery process. The goal definition will implicitly or explicitly define the search space within which we operate. Available structured information can be used to generate or refine a hypothesis within that search space. Often, when we are doing more than pure data analysis, there will be an iterative process of information gathering prior to the final output, or interpretation.}
  \label{fig:our_iterative_process}
  \end{figure}

We note here that this hypothesis-first approach to discovery (Figure~\ref{fig:our_iterative_process}) is more consistent with the philosophy of Popper \cite{popper_conjectures_1962}. This is in contrast to an observation- or experiment-first approach, which is more consistent with the philosophy of Bacon \cite{bacon_novum_1878}; data mining studies tend to be Baconian \cite{giza_automated_2017}. In practice, when discovery proceeds iteratively, the distinction between the two is simply where one enters the cycle. Both are types of model-guided discovery, which is distinct from brute-force screening or approaches relying solely on serendipity that we discuss later.

\section{Elements of autonomous discovery}
It is impossible to imagine conducting research without some degree of machine assistance, defining ``machine'' broadly. We rely on computers to organize, analyze, and visualize data; analytical instruments to queue samples, perform complex measurements, and convert them into structured data. 
However, it is important to consider precisely what is facilitated by automation or computer-assistance in terms of the broader discovery process. Many technologies (e.g., NMR sample changers) add a tremendous amount of convenience and reduce the manual burden of experimentation, but provide only a modest acceleration of discovery rather than a fundamental shift in the way we approach these problems. Considering the cognitive burden of experimental design and analysis connects to the distinction between autonomy and automation.  A toy slot car that sets its own speed as it proceeds through a fixed track is qualitatively different from a self-driving car in the city, yet each successfully operates within its defined environment. Though there is no precise threshold between automation and autonomy, autonomy generally implies some degree of decision-making and adaptability in response to unexpected outcomes.

\subsection{Assessing autonomy in discovery}\label{sec:assessing_autonomy}

Here, we propose a set of questions to ask when evaluating the extent to which a discovery process or workflow is autonomous: (i) How broadly is the goal defined? (ii) How constrained is the search/design space? (iii) How are experiments for validation/feedback selected? (iv) How superior to a brute force search is navigation of the design space? (v) How are experiments for validation/feedback performed? (vi) How are results organized and interpreted? (vii) Does the discovery outcome contribute to broader scientific knowledge? These questions are mapped onto the schematic for hypothesis-driven discovery in \autoref{fig:our_iterative_process_search}.

\begin{enumerate}[(i)]
\item {\bf{How broadly is the goal defined?}} While algorithms can be made to exhibit creativity (e.g., coming up with a unique strategy in Go or Chess \cite{silver_mastering_2017,silver_general_2018}), at some level, they do so for the sake of maximizing a human-defined objective. Is the goal defined at the highest level possible (e.g., find an effective therapeutic)? Or is it narrow (e.g., find a molecule that maximizes this black-box property for which we have an assay and preliminary data)? The higher the level at which the mission can be defined, the more compelling the discovery becomes. That requires platforms to understand what experiments can be performed and how they are useful for the task at hand.

\item {\bf How constrained is the search/design space?} An unconstrained search space is one that \emph{we} operate in as human researchers. There are many ways in which humans can artificially constrain the search space available to an autonomous platform. A maximally constrained search space in the discovery of physical matter could be a (small) fixed list of candidates over which to screen. Limitations in the experimental and computational capabilities of an autonomous platform have the effect of constraining the search space as well; the scientific process itself has been described by some as a dual search in a hypothesis space and experimental space \cite{klahr_dual_1988,klahr_heuristics_1993}. How these constraints are defined influences the difficulty of the search process, the likelihood of success, and the significance of the discovery. The fewer the constraints placed on a platform, the greater the degree to which it can be said to be operating autonomously.  %

\item {\bf How are experiments for validation/feedback selected?} 
Unconstrained experimental design is a complex process requiring evaluation of local decisions as well as a global strategy for the overall timeframe, coherency, and scientific merit of a proposed experiment \cite{baker1996constraints}. When operating within a restricted experimental space, design can be simplified to local decisions of specific implementation details without these high-level decisions. %
\citeauthor{cummings_collaborative_2009} define a taxonomy for human-automation collaboration in terms of the three primary roles played by a human or computer: moderator (of the overall decision-making process), generator (of feasible solutions), and decision-maker (of which action to take) \cite{cummings_collaborative_2009}. Their levels of automation include ones where humans must take all decisions/actions, where the computer narrows down the selection, where the computer executes one if the human approves, and where the computer executes automatically and informs the human if necessary. The second level is typical for the discovery of new physical matter, where computational design algorithms may propose compounds that must be subjected to a manual assessment of synthesizability before being manually synthesized. The smaller the search space and the cheaper the experiments--including considerations of time and risk of failure--the less human intervention is required in selecting experiments.

\item {\bf How superior to a brute force search is navigation of the design space?}\label{item:brute_force_search} This question seeks to identify the extent to which there is ``intelligence'' in the search strategy. \citeauthor{langley_scientific_1987}'s notion of discovery as a heuristic search emphasizes this criterion \cite {langley_scientific_1987}. Whether or not the strategy is more effective than a brute force search depends on the size of the space and how experiments are selected. For example, a high throughput screen of compounds from a fixed library is equivalent to a brute-force search. An active learning strategy designed to promote exploration might require only 20\% of the experiments to find an optimal solution. When dealing with continuous (e.g., process variables) or virtually infinite (e.g., molecular structure) design spaces, it is not possible to quantify meaningfully the number of experiments in a brute-force search. %

\item {\bf How are experiments for validation/feedback performed?} Being able to automatically gather new information to support/refute a hypothesis is an important aspect of an automated discovery workflow. At one extreme, experiments are performed entirely by humans (regardless of how they are proposed); in the middle, experiments might be performed semi-automatically but require significant human set-up between experiments; at the other extreme, experiments can be performed entirely without human intervention. This question is tightly coupled to that of who chooses the experiments and the size of the search space. The narrower the experimental design space, the more likely it is that validation/feedback can be automated. In computational studies, it is relatively straightforward to automate simulations if we are willing to discard failures without manual inspection (e.g., DFT simulations that fail to converge). %

\item {\bf How are results organized and interpreted?} In an iterative workflow, the results of information gathering (experiments, simulations) are organized as structured information and used to update our prior knowledge and revise our beliefs before the next round of experimental design. Provided that the experiments/simulations can be designed to produce information that is already  in a compatible format (e.g., quantifying a reaction yield to build a model of yield as a function of process variables), this is simply a practical step toward closing the loop. In a few specialized workflows, experimental results naturally drive the selection of subsequent experiments, as in directed evolution and phase-assisted continuous evolution \cite{packer_methods_2015}.

\item {\bf (optional) Does the outcome contribute to broader scientific knowledge?} Though not necessarily related to the concept of autonomy, this question speaks to impact and intelligibility. Does it require extensive interpretation after the fact to evaluate \emph{how} or \emph{what} it has learned, or is it self-explanatory? Intelligibility is one of the criteria for discovery put forward by \citeauthor{valdes-perez_principles_1999} \cite{valdes-perez_principles_1999}, among others. Describing physical phenomena requires far less domain knowledge than does explaining those phenomena \cite{langley_data-driven_1989}. Especially in empirical modeling, there is often a dichotomy between models built for accurate predictions and models built for explanatory predictions \cite{breiman_statistical_2001, shmueli_explain_2010}. Turing made note of this at least as early as 1950, saying that ``an important feature of a learning machine is that its teacher will often be very largely ignorant of quite what is going on inside, although he may still be able to some extent to predict his pupil's behavior''  \cite{turing_i.computing_1950}. The past few years have seen an interest in the transparency, interpretability, and explainability of machine learning models, not just the accuracy \cite{roscher2019explainable}. 

\end{enumerate}

Several of these questions probe the extent to which discovery is ``closed loop'', which implicitly assumes an iterative process of multiple hypothesize-test-revise beliefs cycles. Iterative refinement is crucial when operating inside poorly-explored design spaces (e.g., using an uncommon scaffold) or with new objective functions (e.g., maximizing binding to a new protein target \emph{in vitro}).   Most of the case studies described in the following sections are better described as ``open loop'' and involve only certain aspects of the workflow in \autoref{fig:our_iterative_process_search}. For example, a common paradigm of computer-aided discovery is to define an objective function, perform a large-scale data mining study, propose a solution of new molecule, material, and/or model, and manually validate a small number of those predictions. \citeauthor{waltz_automating_2009} describe many early computational discovery programs as merely running calculations, rather than trying to close the loop \cite{waltz_automating_2009}.

\subsection{Enabling factors}

\begin{figure}[h]
  \centering
  \includegraphics[width=6cm]{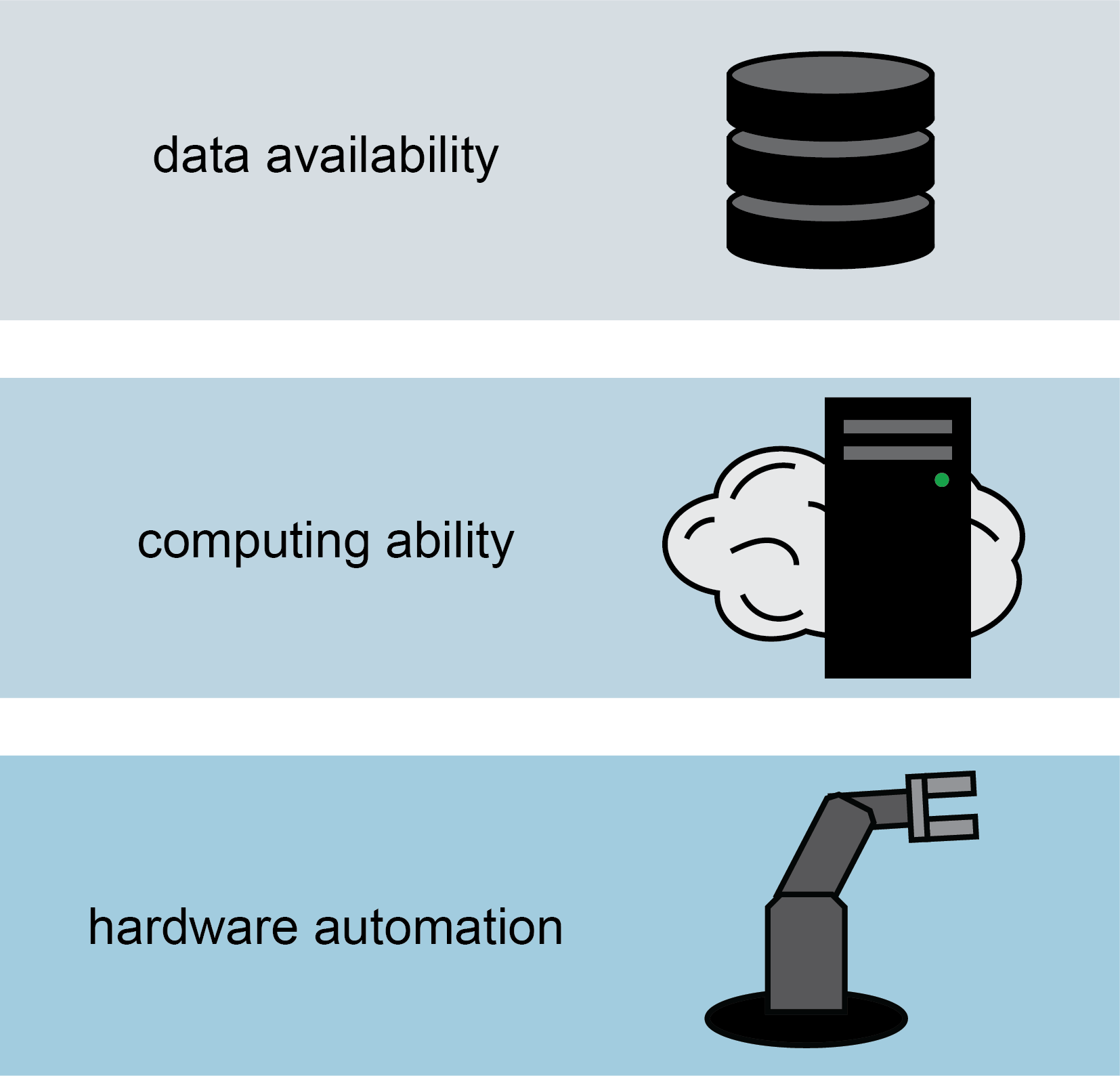}
  \caption{The factors that have enabled autonomous discovery fall into one of three main categories.}
  \label{fig:enabling_pillars}
\end{figure}

A confluence of improved data availability, computing abilities, and experimental capabilities have brought us substantially closer to autonomous discovery (Figure~\ref{fig:enabling_pillars}). These improvements contribute to two categories of methodological progress: (1) techniques for navigating the search space more effectively, and (2) techniques for accelerating validation/feedback. Many machine learning techniques, for example, have been used to build empirical models within the search space to enable or accelerate the search; mapping the design space for a molecule, material, device, or process to relevant performance metrics is a prerequisite for any ``rational design''.

As \citeauthor{claus_discovery_2002} point out, effective discovery requires assimilation of knowledge contained in large quantities of data of diverse types \cite{claus_discovery_2002}. The quantity of chemical property and process data available in journals, the patent literature, and online databases makes it challenging to analyze by hand. Digitization of organic reaction information into computer-readable databases like Reaxys, SPRESI, CASREACT, and Lowe's USPTO dataset has not just facilitated searching that information, but has enabled new analyses thereof \cite{warr_short_2014}. Millions of bioactivity measurements are found in databases like ChEMBL and PubChem, not to mention the dozens of genomic, metabolomic, and proteomic databases that have emerged in the last decade \cite{gaulton_chembl:_2012, rigden_2018_2018}. There are also many repositories for experimental and computational properties of materials, which have facilitated the construction of empirical models to predict new material performance \cite{hill_materials_2016, himanen_data-driven_2019}. \citeauthor{gil_amplify_2014} \cite{gil_amplify_2014} discusses the utility of AI techniques in searching and synthesizing large amounts of information as part of ``discovery informatics'' \cite{claus_discovery_2002,honavar_promise_2014,gil_discovery_nodate}. Even now, an enormous amount of untapped information remains housed in laboratory notebooks and journal articles. For such information to be directly usable, someone must undertake the challenge of compiling the data into an accessible, user-friendly format and overcome any intellectual property restrictions. Image and natural language processing techniques can make this task less burdensome; there is increasing interest in adapting such information extraction algorithms for use in chemistry \cite{krallinger2015chemdner,swain2016chemdataextractor, krallinger_information_2017, kim_materials_2017, zhai_improving_2019, zheng_text_2019}.

Autonomous discovery systems rely on a variety of computational tools to generate hypotheses from data without human intervention. This includes both the software that makes the recommendations (e.g., proposes correlations, regresses models, selects experiments) as well as the underlying hardware that makes using the software tractable. Our discussion of the advances in this area focuses on software developments with an emphasis on machine learning algorithms, which has elicited cross-disciplinary excitement \cite{musib_artificial_2017, chen_rise_2018, goh_deep_2017, vamathevan_applications_2019}. 

Typically, search domains that are of interest for discovery are characterized by high dimensionality (e.g., chemical space). In such domains, the patterns within the available data may be beyond the capacity of humans to infer {\it{a priori}} without years of intuition-building practice. Machine learning and pattern recognition algorithms can be used to discover these regularities automatically, e.g., by using the available data to parameterize a neural network model \cite{Bishop2006}. \citeauthor{Varnek2012} and \citeauthor{Mitchell2014} provide overviews of machine learning techniques as applicable to common cheminformatics problems \cite{Varnek2012, Mitchell2014} and brief tutorials can be found in a number of reviews \cite{luts2010tutorial,rupp2015machine,mueller2016machine,butler_machine_2018}. It is becoming increasingly common to use machine learning to develop empirical quantitative structure-activity/property relationships (QSARs/QSPRs) to score molecules and guide virtual screening as part of broader discovery frameworks \cite{zhang_machine_2017}. These models can be used to distinguish promising compounds from unpromising ones and prioritize molecules for synthesis and testing (validation), thus facilitating the extrapolation of information about existing molecules to novel molecules that exist only {\it{in silico}} \cite{Walters2018}.

Algorithms that enable efficient navigation of design spaces represent an important set of computing advances. Even with a  model representing our belief about a physical structure-property relationship, an algorithmic framework is needed to apply that belief to experimental design. These frameworks include active learning strategies \cite{settles2012active} that aim to maximize the accuracy of predictive models while minimizing the required training data, as well as goal-directed strategies such as Bayesian optimization \cite{frazier2018tutorial} and genetic algorithms \cite{holland1992adaptation}. These iterative techniques can reduce the experimental burden associated with discovery in domains or search spaces where exhaustive testing is not practical. 

\begin{figure}[h]
  \centering
  \includegraphics{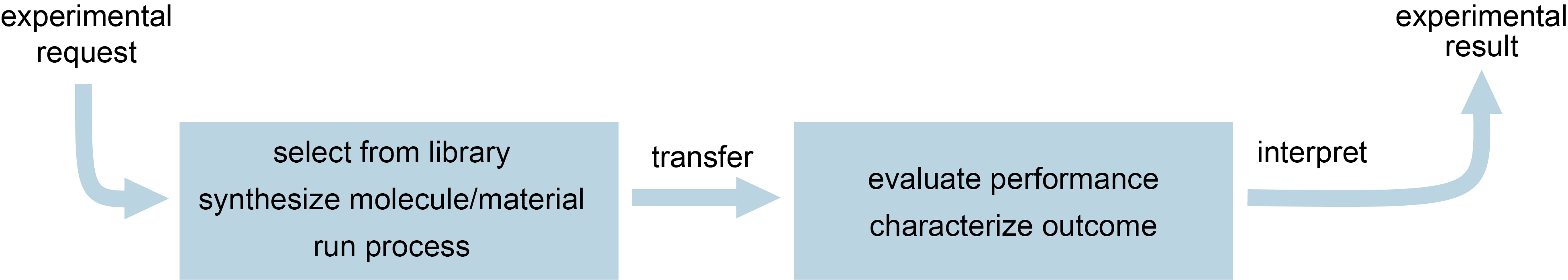}
  \caption{Generic workflow for experimental validation.}
  \label{fig:experimental_systems}
\end{figure}

Algorithms that are capable of directly proposing candidate molecules or materials (physical matter) as a form of experiment selection are worth special emphasis. Recently, deep generative models \cite{Salakhutdinov2015} such as generative adversarial networks (GANs) \cite{goodfellow2014generative} and variational autoencoders (VAEs) \cite{Kingma2014} have attracted a great deal of interest, as they facilitate the creation of diverse molecular libraries without the impossible task of systematically enumerating all potential functional compounds \cite{sanchez-lengeling_inverse_2018, schwalbe-koda_generative_2019, elton_deep_2019}. Many case studies that leverage these and related frameworks for the discovery of physical matter are described later.
Experimental advances toward autonomous discovery include automation of well-established laboratory workflows (along with parallelization and miniaturization) as well as entirely novel synthetic and analytical methodologies. Aspects of experimental validation (\autoref{fig:experimental_systems}) have existed in an automated format for decades (e.g., addition to and sampling from chemical reactors \cite{deming_automated_1971, winicov_chemical_1978}), and many of the requisite hardware units have been commercialized (e.g., liquid handling platforms and plate readers available through companies such as Beckman, Hamilton, BioTek, and Tecan). However, moving beyond piecemeal automation to the entire experimental burden of discovery workflows is challenging. Each process step, which may include synthesis, purification, assay preparation, and analysis, must be seamlessly integrated for the platform to operate without manual intervention; each interface presents new potential points of failure \cite{Pan2019}. 

The complexity of software required for hardware automation ranges from sequencing commands from a fixed schedule \cite{steiner_organic_2018}, to real-time control and optimization \cite{fitzpatrick_novel_2016}, to higher-level scheduling and orchestration \cite{roch_chemos:_2018}; user interface driven software such as LabVIEW \cite{elliott_national_2007} can  aid the creation of fit-for-purpose control systems with minimal programming experience. Although end-to-end automation of an experimental discovery workflow is uncommon, there are numerous benefits to be gained even from partial automation, chief among these being standardization and increased throughput \cite{chapman_lab_2003}. 

In addition to automation, novel experimental methodologies have been developed that lend themselves particularly well to autonomous discovery workflows by  facilitating the exploration of broad design spaces, a helpful feature that increases the likelihood of discovery \cite{Kodadek2010}. These include synthesis-focused methodologies, such as DNA-encoded libraries \cite{brenner_encoded_1992} and diversity-oriented synthesis \cite{schreiber_target-oriented_2000, Tan2005}, as well as analysis-focused methodologies, such as ambient mass spectrometry \cite{cooks_ambient_2006} and MISER for accelerating liquid chromatographic analysis \cite{welch2010miser}.

The three categories of enabling factors described herein facilitate discovery in different ways: data is leveraged to create models that inform and predict, computational tools are used to create models from data and reason about which experiments to perform next, and physical (or computational) experiments validate hypotheses and facilitate refinement thereof. These factors can be strategically combined to give rise to different types of studies. For example, the experimental capabilities described here, in isolation, can be used for high-throughput, brute-force screening; computational tools can be used for data generation (through, e.g., DFT simulations); virtual screening is achieved through the combination of data and algorithms; and integration of all three is needed for fully autonomous discovery.

\iffull %

\section{Examples of (partially) autonomous discovery}
\label{sec:cases}

In this section, we summarize a series of case studies that demonstrate how automation and machine autonomy influence discovery in various research domains. 
The extent to which techniques in automation and computation have enabled each case varies. Some only benefit from automated laboratory hardware, others learn underlying trends from large or complex data, and still others use computational techniques to efficiently explore high dimensional design spaces. %

Specifically, subsection \ref{sec:cases:early} describes early computational reasoning frameworks;  \ref{sec:cases:mechanistic} describes the discovery of mechanistic models; \ref{sec:cases:noniterative_process} and \ref{sec:cases:iterative_process} describe the noniterative and iterative discovery of chemical processes; \ref{sec:cases:noniterative_qsar} describes the noniterative discovery of property models; \ref{sec:cases:noniterative_phys} and \ref{sec:cases:iterative_phys} describe the noniterative and iterative discovery of physical matter; finally, \ref{sec:cases:brief_summary} provides a brief summary of a few tangentially-related domains.

\subsection{Foundational computational reasoning frameworks}
\label{sec:cases:early}

There has been a long-standing fascination with the philosophical question of whether or not it is possible to codify and automate the process of discovery \cite{simon_scientific_1981}. In the 1980s and 1990s, several programs were developed to mimic a codifiable approach to discovery and to reproduce specific quintessential discoveries \emph{of models}, led by Langley and Zytkow \cite{sparkes_towards_2010}. These programs deal with questions of model induction and hypothesis generation (as a form of data analysis) rather than experimental selection and automated validation/feedback.

{BACON} is a rule-based framework introduced in 1978 to formalize the Baconian method of inductive reasoning to discover empirical laws, supplemented with data-driven heuristics \cite{langley_bacon._1978}. {BACON.4}, a later iteration specifically designed for chemical problems, searched for arithmetic combinations of input variables to identify regularities in data (e.g., noting that pressure times volume is invariant for constant temperature in a closed gas system) \cite{sozou_computational_2017}. This approach was able to recapitulate Ohm's law, Archimedes' law of displacement, Snell's law, conservation of momentum, gravitation, and Black's specific heat law  \cite{langley_rediscovering_1983}. The search for an empirical relationship was greatly simplified by excluding any irrelevant variables (i.e., all input variables were known to be important) and eliminating all measurement noise. Extensions of this approach included describing piecewise functions ({FARENHEIT} \cite{zytkow_combining_1987}) and coping with irrelevant observations and noise ({ABACUS} \cite{falkenhainer_integrating_1986}). More recently, \citeauthor{schmidt_distilling_2009} demonstrated that using a symbolic regression framework similar to BACON, it is possible to rediscover Hamiltonians, Lagrangians, and geometric conservation laws from empirical motion tracking data \cite{schmidt_distilling_2009}. Much like its predecessors, their program uses a two-part process of generating and scoring hypothesized analytical laws.

The STAHL program developed by \citeauthor{zytkow_theory_1986} in the mid-1980s sought to automate the construction of compositional models to, e.g., rediscover Lavoisier's theory of oxygen \cite{zytkow_theory_1986}. It operates on a list of chemical reactions to produce a list of proposed chemical elements and the compounds they make up by making inferences like ``\ce{A + B + C -> B + D}'' $\implies$ ``D is composed of A and C''. While the program was arguably successful in formalizing a specific form of scientific reasoning, the lack of any consideration for stoichometry, phase changes, and ability to consider uncertainty, competing hypotheses, and request information makes such a logic framework highly limited in utility. The KEKADA program \cite{kulkarni_processes_1988} was designed with those abilities in order to replicate the discovery of the Krebs cycle. Using seven heuristic operators (hypothesis proposers, problem generators, problem choosers, expectation setters, hypothesis generators, hypothesis modifiers, and confidence modifiers) and  simulated experiments of metabolic reactions, KEKADA was able to rediscover the Krebs cycle from the same empirical data that would have been obtainable at the time.

The knowledge bases for these early programs were comprised of expert-defined relationships, rules, and heuristics designed to reflect {\it prior knowledge} and bring the programs up to the level of domain experts. Programs based entirely on user-defined axioms have proved successful in automatic theorem generation in graph theory \cite{fajtlowicz_conjectures_1988}. However, these rules bring at least two drawbacks in the context of inductive reasoning. The first is that it is more difficult for experts to recapitulate their knowledge through rules than by providing examples from which an algorithm can generalize \cite{witten_using_1988}.  The second is that too stringent priors may restrict the model from deviating far enough from existing theory to make a substantial discovery and  merely ``fill in the gaps'' of what is known. \citeauthor{kulkarni_processes_1988} argue that a lack of prior knowledge about allowed/disallowed reactions actually served to \emph{benefit} Krebs, as a formally trained chemist might not have pursued a hypothesis that was--at the time--believed to be highly unlikely \cite{kulkarni_processes_1988}.

\subsection{Discovery of mechanistic models}
\label{sec:cases:mechanistic}

\subsubsection{Discovery of detailed kinetic mechanisms}

Computer assistance has proved useful in the exploration and simulation of reaction pathways \cite{green_predictive_2007,simm_error-controlled_2018,simm_exploration_2019, unsleber_exploration_2019}. The vast number of possible elementary reactions creates a combinatorial space of hypothesized pathways that is difficult to explore manually in an unbiased manner, making it a prime candidate for algorithmic approaches. One such approach, MECHEM, enumerates elementary reactions in catalytic reaction systems to identify series of mechanistic steps able to rationalize an observed global reaction \cite{valdes-perez_human/computer_1994,valdes-perez_conjecturing_1994,valdes-perez_machine_1995}.  \citeauthor{ismail_automatic_2019} have demonstrated a similar approach to identifying multi-step reaction mechanisms for catalytic reactions using a ReaxFF potential energy surface \cite{senftle_reaxff_2016} to guide the search toward kinetically-likely pathways \cite{ismail_automatic_2019}. In the absence of heuristics or calculations to drive the search, millions of possible elementary reactions can be generated even with species of just a few atoms \cite{margraf_systematic_2019}. %

The Reaction Mechanism Generator (RMG) fills a similar role in developing detailed kinetic mechanisms for combustion and pyrolysis processes \cite{gao_reaction_2016}. Expert-defined reaction templates  enumerate potential elementary reactions between a set of user-defined input molecules; rate constants for the forward and reverse reactions are estimated from a combination of first principles calculations (e.g., DFT) and group additivity rules regressed to experimental data. %
The ability to estimate kinetic and thermodynamic parameters enables the identification of new elementary reactions and pathways and, e.g., exploration of untested fuel additives' effects on ignition delay \cite{zhang_modeling_2018}. An earlier study by \citeauthor{broadbelt_computer_1994} used a similar approach to develop detailed kinetic models for pyrolysis reactions \cite{broadbelt_computer_1994}.

\begin{figure}[h]
  \centering
  \includegraphics[width=8cm]{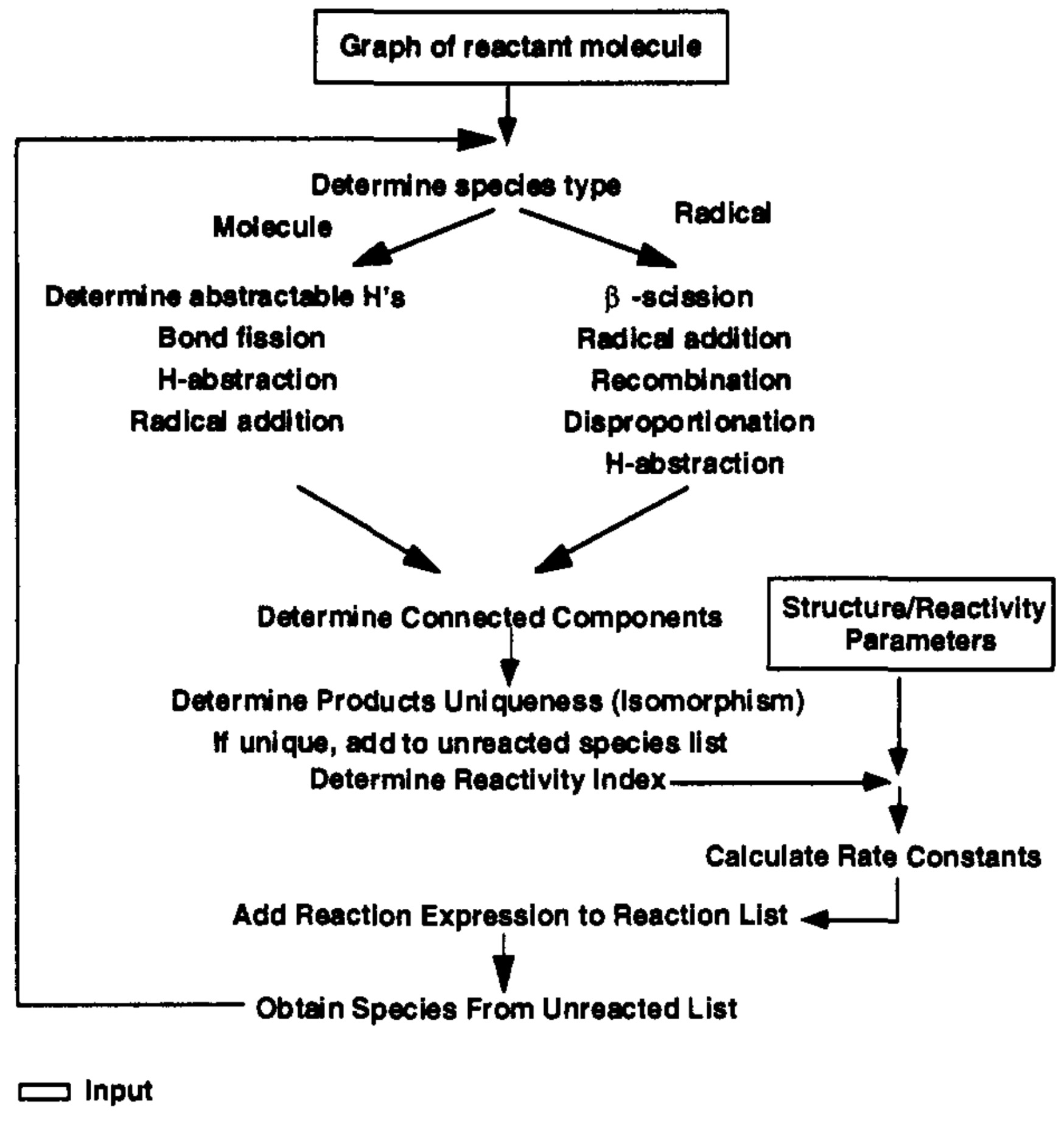}
  \caption{Discovery of detailed kinetic models through iterative selection of important elementary reaction steps. Figure reproduced from \citeauthor{broadbelt_computer_1994} \cite{broadbelt_computer_1994}.}
  \label{fig:broadbelt_1994_mechanism_search}
\end{figure}

Mechanistic enumerations/searches have been applied extensively to the discovery of transition states and reaction channels \cite{zimmerman_automated_2013, ulissi_address_2017, maeda_benchmarking_2019}. These methods represent a search in the $(3N-6)$-dimensional potential energy surface landscape implicitly defined by an $N$-atom pool of reacting species. Approaches like Berny optimization \cite{schlegel_optimization_1982} are used to identify transition state (TS) geometries for the purposes of estimating energetic barrier heights. Double-ended search methods like the freezing string method (FSM \cite{behn_efficient_2011}) or growing string method (GSM \cite{goodrow_transition_2009}) require knowledge of the product structure and run iterative electronic structure calculations to identify a plausible reaction pathway; these can be applied to the discovery of new elementary reactions by systematically enumerating potential product species \cite{suleimanov_automated_2015,grambow_unimolecular_2018,kim_efficient_2018} (Figure~\ref{fig:kim_efficient_2018}). Single-ended search methods operate on reactant species only and perturb the geometry along reactive coordinates, including, e.g., the artificial force induced reaction method (AFIR \cite{maeda_exploring_2014}). 
An alternate approach to reaction discovery is by direct simulation of reactive mixtures using molecular dynamics (MD) \cite{wang_discovering_2014, wang_automated_2016, lei_mechanism_2019}. \citeauthor{wang_discovering_2014} describe the use of an ``\emph{ab initio} nanoreactor'' to find unexpected products from similar starting materials to the Urey-Miller experiment on the origin of life \cite{wang_discovering_2014}. Importantly, their approach does not require the use of heuristics to define reaction coordinates or enumeration rules to define possible products. Instead, molecules in an MD simulation are periodically pushed toward the center to impart kinetic energy and encourage collisions at a rate that enables the observation of rare events over tractable simulation timescales. In principle, these can be to applied to the prospective prediction of novel reaction types and, ultimately, the development of new synthetic methodologies. 

\begin{figure}[h]
  \centering
  \includegraphics[width=\onecolumnsize]{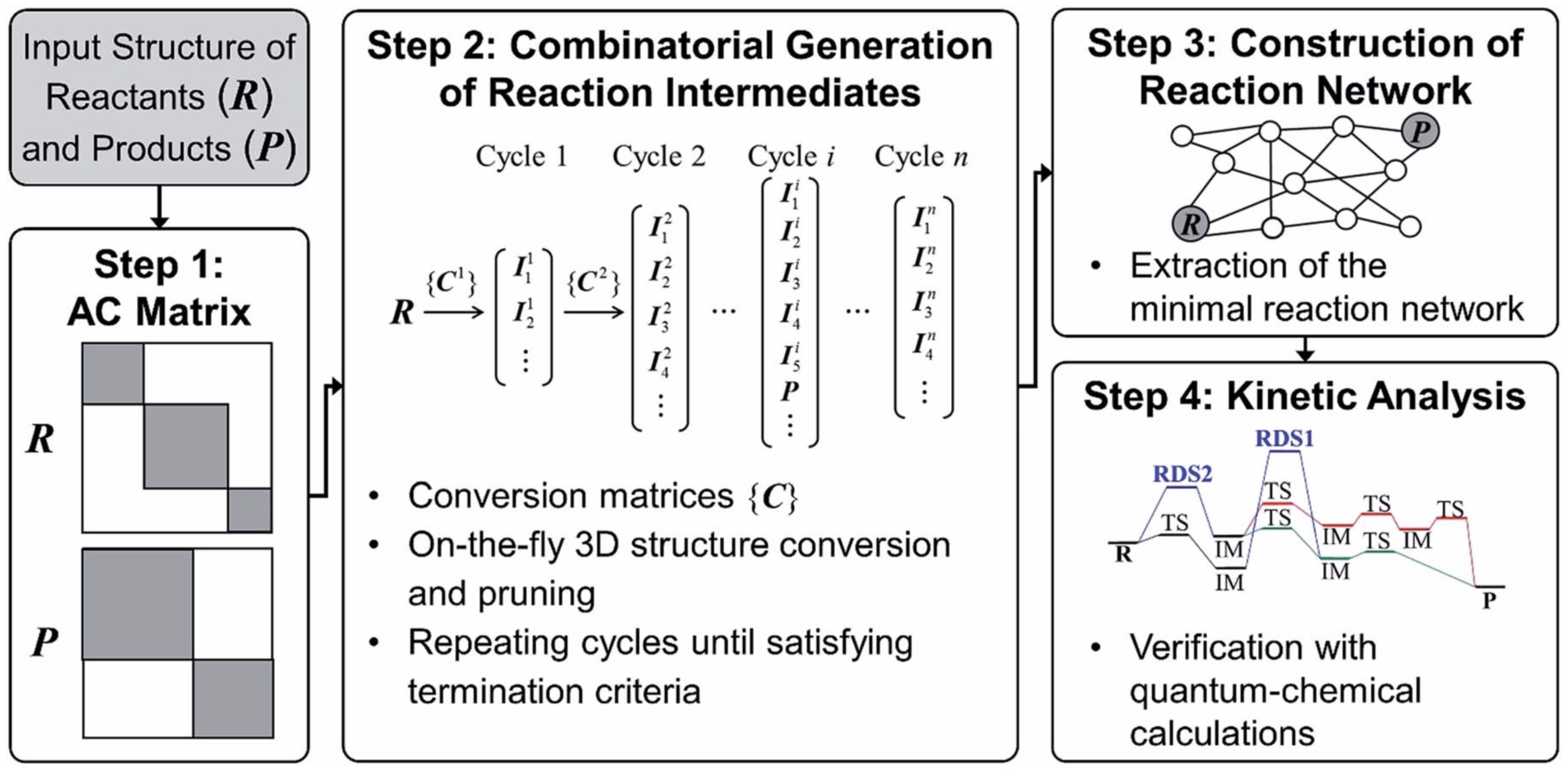}
  \caption{Workflow for identification of reaction networks between known reactants R and known products P through combinatorial enumeration of possible mechanistic steps pruned by calculated transition state energies. Figure reproduced from \citeauthor{kim_efficient_2018} \cite{kim_efficient_2018}.}
  \label{fig:kim_efficient_2018}
\end{figure}

\subsection{\emph{Noniterative} discovery of chemical processes} 
\label{sec:cases:noniterative_process}

\subsubsection{Discovery of new synthetic pathways} %

Synthetic pathways are a prerequisite for physically producing a molecule of interest, whether for experimental validation of a predicted property or for production at scale.  Retrospective analyses of known single-step chemical reactions can yield hypothesized synthetic pathways as combinations thereof.
\citeauthor{gothard_rewiring_nodate} describe an analysis of a ``Network of Organic Chemistry''--a copy of the Beilstein database with seven million reactions--for the discovery of one-pot reactions; their search space comprised any consecutive sequence of known reactions where the product of one is a reactant of another \cite{gothard_rewiring_nodate}. Candidate sequences were evaluated using eight filters, including a $322\times 322$ table of functional groups and their cross-reactivity and a $322\times 97$ table of their compatibility under 97 categories of reaction conditions. 
Through application of these expert heuristics to millions of candidate sequences, the authors identified multi-step chemistries that could potentially be run without an intermediate purification, choosing a handful of such pathways for experimental validation.  While their filters were all hand-encoded, data mining techniques can also be used to estimate functional group reactivity \cite{soh_estimating_2012,lin_automatized_2016}.  Selecting pathways within a search space defined by combinations of known single-step reactions has taken on other forms as well, including the identification of cyclic pathways \cite{bajczyk_discovery_2018}, the optimization of process cost \cite{lapkin_automation_2017}, and the optimization of estimated process mass intensity \cite{li_making_2019}.

Generating yet-unseen chemical reactions for a synthesis plan--a necessity for the synthesis of novel molecules--is a harder search problem than when searching within a fixed reaction network \cite{smith_computational_1997}. Because the number of states in a naive retrosynthetic expansion will scale as $b^d$ for branching factor $b$ and depth $d$, guiding the search is an essential aspect of computer-aided synthesis planning (CASP) programs. The breadth of the search depends on the coverage of the rule sets: abstracted enzymatic reactions tend to number in the hundreds \cite{kim_review_2017}, expert transformation rules often number in dozens or hundreds \cite{Hartenfeller2012, avramova_retrotransformdb:_2018} but can extend into the tens of thousands in contemporary programs \cite{szymkuc_computer-assisted_2016}, and algorithmically-extracted templates generally number in the thousands to hundreds of thousands \cite{law_route_2009, christ_mining_2012, bogevig_route_2015, segler_neural-symbolic_2017}. To the extent that reaction rules and synthetic strategies can be codified, synthesis planning is highly conducive to computational assistance \cite{ihlenfeldt_computer-assisted_1996, h.todd_computer-aided_2005, cook_computer-aided_2012, ravitz_data-driven_2013, warr_short_2014, coley_machine_2018} (Figure~\ref{fig:ihlenfeldt_computer-assisted_1996}). CASP approaches that generate retrosynthetic suggestions without the use of pre-extracted template libraries \cite{liu_retrosynthetic_2017,zheng_predicting_2019} still result in a large search space of possible disconnections.

Even the earliest CASP programs emphasized the importance of navigating the search space of possible disconnections \cite{corey_computer-assisted_1969, ihlenfeldt_computer-assisted_1996}. The search in OCSS was guided by five subgoals for structural simplification: reduce internal connectivity, reduce molecular size, minimize functional groups, remove reactive or unstable functional groups, and simplify stereochemistry \cite{corey_computer-assisted_1969}. Starting material oriented retrosynthesis introduces additional constraints in the search, as the goal state is a specific starting material, rather than one of many from a database of available compounds \cite{johnson_recent_1992}. It is only fairly recently that CASP tools have started to be used more widely for discovery of synthetic routes. Development is stymied by the complexities of validation and feedback, which can only occur by experimental implementation \cite{klucznik_efficient_2018} or review by expert chemists \cite{segler_planning_2018}.

\begin{figure}[h]
  \centering
  \includegraphics[width=\onecolumnsize]{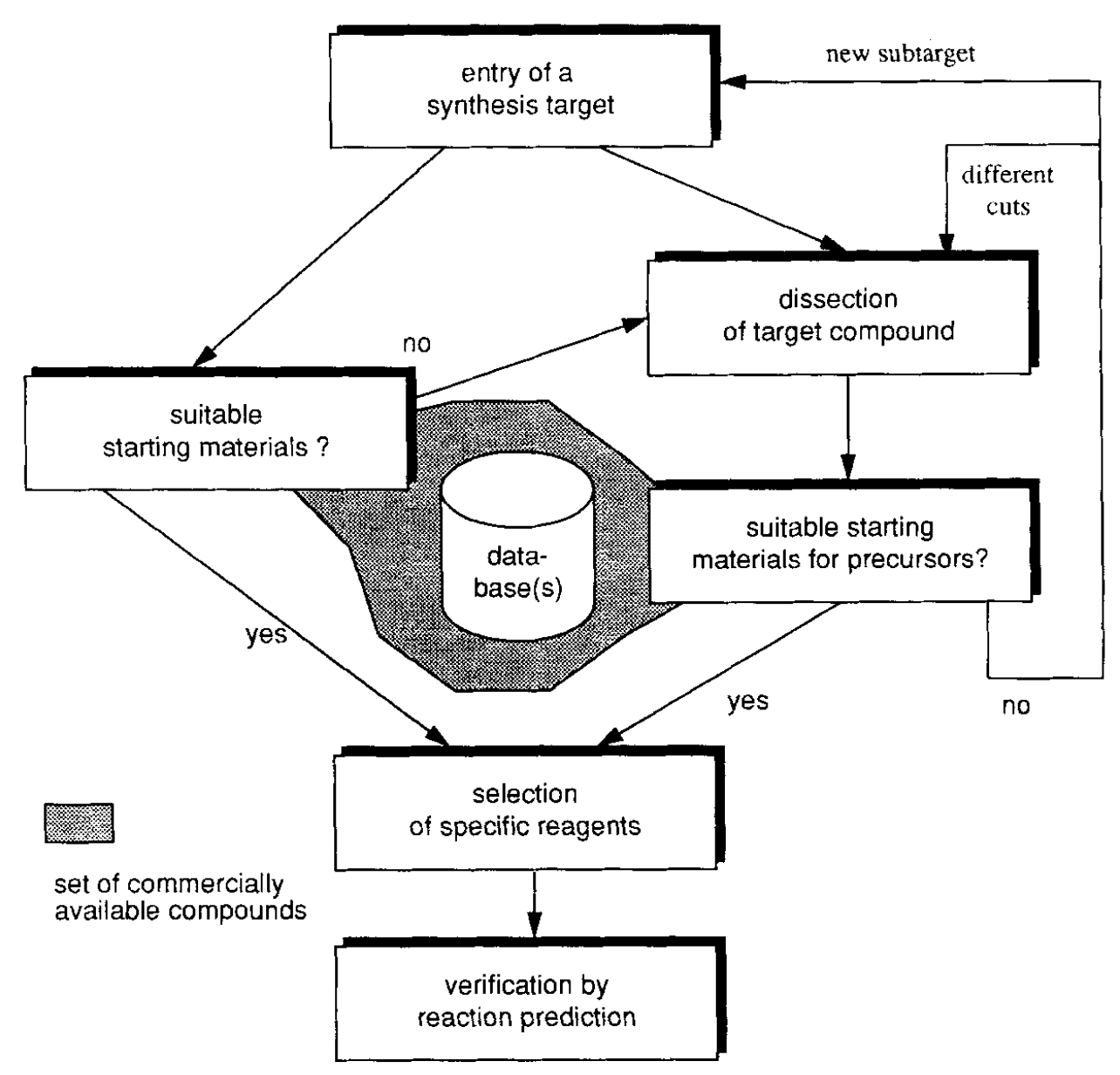}
  \caption{Workflow used by the WODCA program for computer-aided synthesis planning. Figure reproduced from \citeauthor{ihlenfeldt_computer-assisted_1996} \cite{ihlenfeldt_computer-assisted_1996}.}
  \label{fig:ihlenfeldt_computer-assisted_1996}
\end{figure}

There are two main approaches to navigating the search space during retrosynthetic expansion to determine which disconnections are most promising: value functions and action policies. Value functions  estimate the synthetic complexity of reactant molecules as a proxy for how close they are to being purchasable \cite{bertz_first_1981, ertl_estimation_2009, sheridan_modeling_2014, proudfoot_molecular_2017, coley_scscore:_2018}. Despite their limitations, these are widely used in virtual screening libraries as a rapid means of prioritizing compounds that appear more synthetically tractable. While even simple user-defined heuristics that attempt to break a molecule into the smallest possible fragments can be successful in planning full synthetic routes, learned value functions can offer some advantages in finding shorter pathways or being tailored to a user-defined cost function \cite{schreck_learning_2019}. Action policies directly predict which transformation rule to apply based on literature precedents in a knowledge base; this can be accomplished through a simple nearest-neighbor strategy \cite{coley_computer-assisted_2017} or through a trained neural network model for classification \cite{segler_neural-symbolic_2017}. The latter approach has been integrated into a Monte Carlo tree search framework to rapidly generate and explore the space of candidate pathways, resulting in recommendations that chemists considered equally plausible to literature pathways in a double-blind study \cite{segler_planning_2018}. Less common approaches to navigating the search space include proof-number search \cite{heifets_automated_2014}.

Reaction pathway discovery is relevant in synthetic biology and metabolic engineering contexts as well. For example, one study by \citeauthor{rangarajan_language-oriented_2012} describes the application of Rule Input Network Generator (RING,  \cite{rangarajan_rule-based_2010}) to identify plausible production biosynthetic pathways through a heuristic-driven network generation and analysis \cite{rangarajan_language-oriented_2012}. \citeauthor{kim_review_2017} review algorithms and heuristics used to explore metabolic networks and find optimal pathways \cite{kim_review_2017}. A broader review of machine learning for biological networks can be found in ref.~\citenum{camacho_next-generation_2018}.

\subsubsection{Discovering models of chemical reactivity}
Identifying synthetic pathways is but one step toward fully automated synthesis.  For any theoretical robo-chemist capable of synthesizing any molecule on demand \cite{peplow_organic_2014,lowe_derek_automated_2018}, these ideas must be able to be acted upon and executed in the laboratory. Even without automated synthesis, hypothesized synthetic pathways are of little use without experimental validation. This requires additional models of chemical reactivity that can, among other things, propose suitable reaction conditions, estimate the confidence in the reactions it proposes, and have some notion of why one set of substrates might achieve a higher yield than others. Models for these tasks can be trained directly on experimental data using a variety of statistical techniques.

Given a set of combination of successful and unsuccessful reaction examples (i.e., high and low yielding), one can train a binary classifier model to predict whether a proposed set of reaction conditions will be successful \cite{marcou_expert_2015}. The same task can also be treated as a regression of reaction yields, rather than as a classification, as a function of substrate descriptors; a virtual screen of known conditions as a fixed search space can then propose substrate-dependent optimal conditions  \cite{nielsen_deoxyfluorination_2018}. When \emph{only} successful reaction examples are present, one can treat the selection of reaction conditions as a recommendation problem comprising a classification subproblem (for reagent, catalyst, solvent identity) and a regression subproblem (temperature) under the assumption that the ``true'' published conditions are adequate. This was \citeauthor{gao_using_2018}'s approach using the Reaxys database to produce a model that is able to propose conditions at the level of species identity and temperature based on reactant and product structures \cite{gao_using_2018}. In the process of learning the relationship between reactants/products and suitable reaction conditions, the model learns a continuous embedding for chemicals that reflects their function in organic synthesis, similar to how semantic meaning is captured by word2vec models \cite{mikolov_efficient_2013}. Formulating condition selection as a data-driven classification problem has also been used in a more focused manner as an alternative to expert recommender systems \cite{banares-alcantara_decadehybrid_1988}, e.g., to choose phosphine ligands for Buchwald-Hartwig aminations \cite{li_making_2019} or catalysts for deprotections \cite{lin_automatized_2016}.

In some cases, computational prediction of solvation free energies can meaningfully assist in the selection of reaction solvents \cite{reichardt_solvents_2011}. To a first approximation, solvation energy can be estimated by a linear model describing potential solute-solvent interactions \cite{wells_linear_1963,taft_linear_1985}. When those interaction parameters can be predicted via DFT, one can estimate the performance of a large virtual set of solvents, e.g., to optimize the rate constant for a particular reaction of interest \cite{struebing_computer-aided_2013}.
Similar models for \emph{a priori} evaluation of reaction conditions can be found in materials applications. In one instance, \citeauthor{raccuglia_machine-learning-assisted_2016} used a combination of 3955 reaction successes and failures from laboratory notebooks to train an SVM model to predict outcomes for the crystallization of vanadium selenites \cite{raccuglia_machine-learning-assisted_2016} (Figure~\ref{fig:raccuglia_machine-learning-assisted_2016}). Recasting the model as a decision tree led to correlations that reflected expert intuition, which arguably contributed to the synthesis of five previously-unseen compounds \cite{j.xu_understanding_2018}. A similar study applied a much smaller dataset of 54 conditions to predict whether a process would produce atomically precise gold nanocrystals, using a siamese neural network architecture to relate proposed conditions to precedents \cite{li_deep_2018}. For larger scale analyses, the literature serves as an unstructured data source  of inorganic reactions and has been used to populate a structured database of synthesis conditions and outcomes via natural language processing of over 640,000 manuscripts \cite{kim_machine-learned_2017}; virtual screening and synthesis planning pipelines have been built on top of such data to help guide the experimental realization of computationally-proposed materials \cite{kim_materials_2017,kim_virtual_2017,kim_inorganic_2018,jensen_machine_2019}.

\begin{figure}[h]
  \centering
  \includegraphics[width=\onecolumnsize]{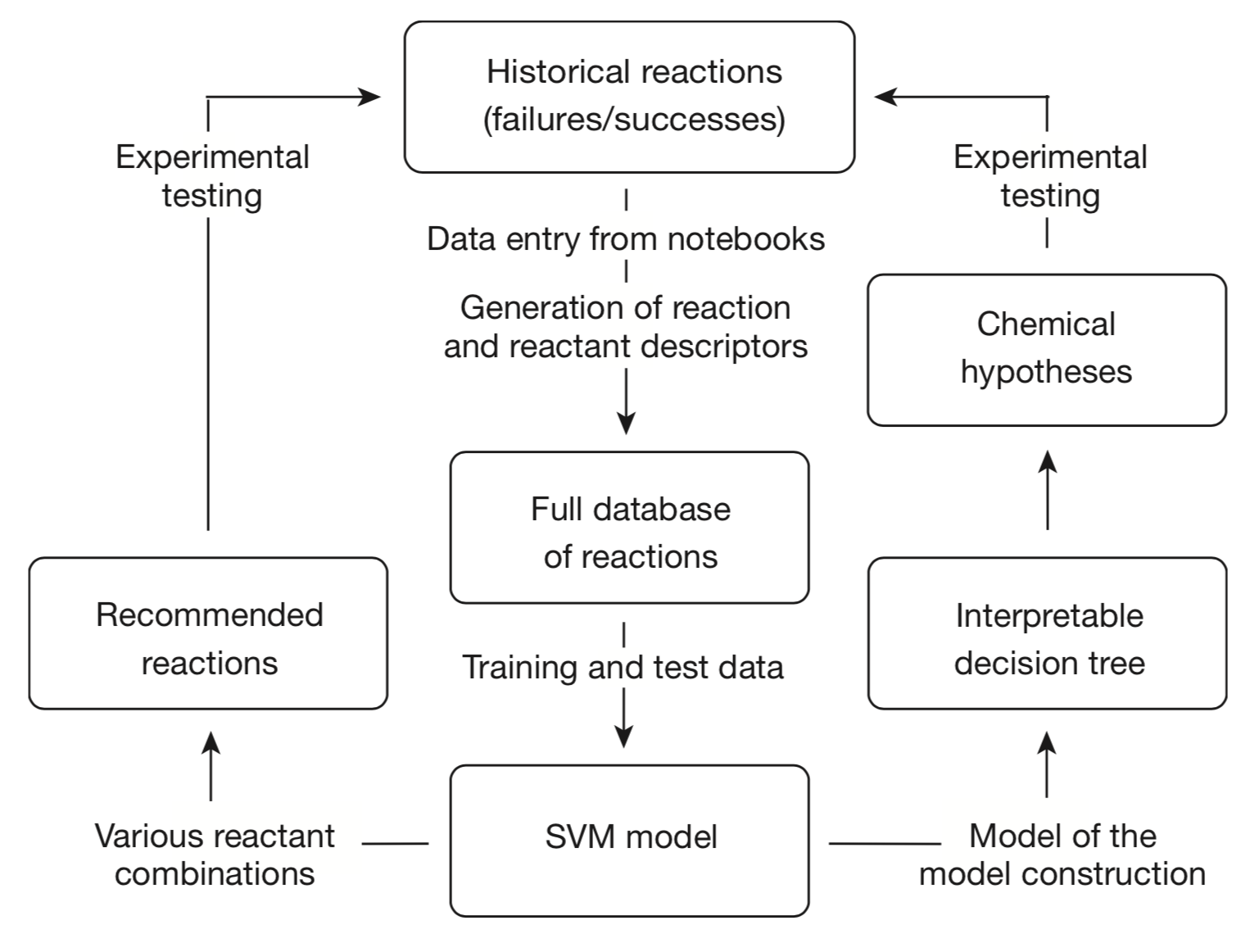}
  \caption{Workflow used by \citeauthor{raccuglia_machine-learning-assisted_2016} for training an interpretable predictive model of the success/failure of vanadium selenite crystallization. Figure reproduced from \citeauthor{raccuglia_machine-learning-assisted_2016} \cite{raccuglia_machine-learning-assisted_2016}.}
  \label{fig:raccuglia_machine-learning-assisted_2016}
\end{figure}

Anticipating the outcomes of organic reactions is a very different modelling task. The space of possible results is high dimensional (chemical space) rather than low dimensional (e.g., the phase of the resulting material or a boolean measure of success/failure). The ability to accurately prediction reaction products would be powerful in combination with CASP to improve the likelihood that proposed reactions are experimentally realizable. The task of predicting reaction outcomes \emph{in silico} has been approached through several heuristic and computational techniques over the years \cite{gasteiger_eros_1978, ugi_new_1979, salatin_computer-assisted_1980, sello_reaction_1992, satoh_sophia_1995, socorro_robia:_2005} but has seen renewed interest as a supervised learning problem as a result of increased data availability \cite{coley_machine_2018}. 

\citeauthor{segler_modelling_2017} treat reaction discovery as an edge prediction problem in a knowledge graph of known chemistry \cite{segler_modelling_2017}. Specifically, they predict the products of bimolecular reactions through the application of algorithmically-extracted half reactions that similar substrates underwent. Novel combinations of half reactions that had not been observed previous could be accurately predicted, albeit with a modest rate of success.  With a similar goal, \citeauthor{jacob_prediction_2018} build a stochastic block model (SBM) for the classification of reactions into true or false using reactions in Reaxys (true) and ones randomly generated from known chemicals (false) \cite{jacob_prediction_2018}.  Other machine learning-based methods include ones that rank enumerated mechanistic \cite{kayala_learning_2011,kayala_reactionpredictor:_2012,fooshee_deep_2017} or pseudo-mechanistic \cite{bradshaw_predicting_2018} steps, score/rank reaction templates \cite{wei_neural_2016, segler_neural-symbolic_2017}, score/rank candidate products generated from reaction templates \cite{coley_prediction_2017}, propose reaction products as resulting from sets of graph edits \cite{jin_predicting_2017,coley_graph-convolutional_2019}, and translate reactant SMILES strings to product SMILES strings using models built for natural language processing tasks \cite{nam_linking_2016,schwaller_found_2017,schwaller_molecular_2018}. These all formulate reaction prediction differently; for example, the model in ref.~\citenum{coley_graph-convolutional_2019} learns to enumerate likely changes in bond order and learns to rank candidate products generated through combinatorial enumeration of those sub-reactions (Figure~\ref{fig:coley_graph-convolutional_2019}).

\begin{figure}[h]
  \centering
  \includegraphics[width=14cm]{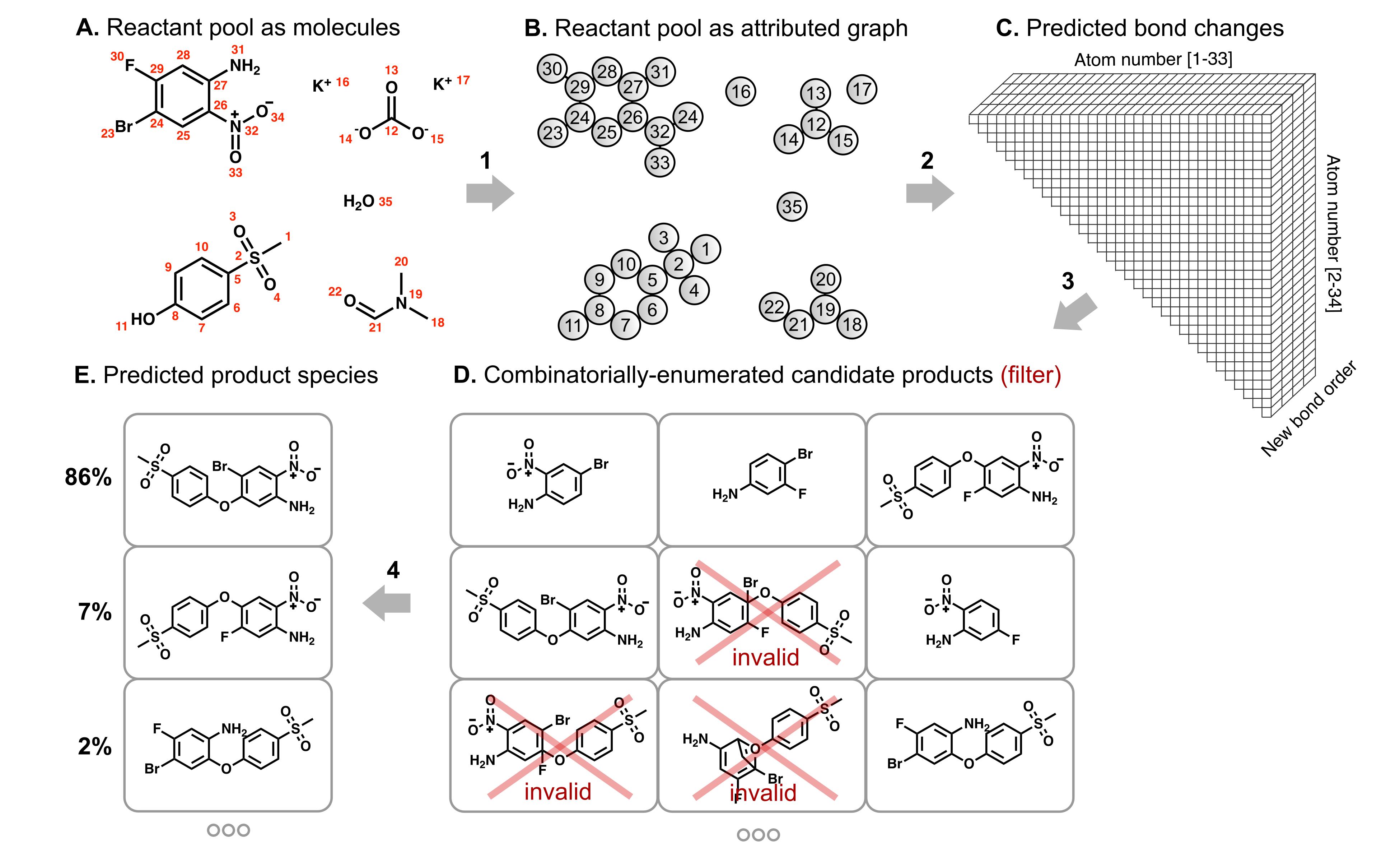}
  \caption{Workflow used by \citeauthor{coley_graph-convolutional_2019} for predicting the products of organic reactions. Figure reproduced from \citeauthor{coley_graph-convolutional_2019} \cite{coley_graph-convolutional_2019}.}
  \label{fig:coley_graph-convolutional_2019}
\end{figure}

The \emph{quantitative} prediction of reaction outcomes is closer to a standard regression task; when only one chemical species is varied--a single substrate or a single catalyst--the problem is exactly that of developing a QSAR/QSPR model. The historical approach in physical organic chemistry is again the development of linear free energy relationships \cite{wells_linear_1963}, for which group contribution approaches are particularly attractive due to their interpretability \cite{platts_estimation_1999}. Hammett parameters are a classic example of correlating molecular structures with reactivity \cite{hammett_effect_1937}. Computational prediction of organic reaction rates has been demonstrated using simple regressions on expert descriptors \cite{gasteiger_computer-assisted_1995} and using structure-derived descriptors \cite{galvez_application_2010, madzhidov_structure-reactivity_2014, polishchuk_structurereactivity_2017, glavatskikh_predictive_2019, madzhidov_structurereactivity_2017} with much of the latter work coming from Varnek and coworkers. 

Even with increasingly powerful machine learning techniques to describe patterns in experimental data, computational chemistry has a significant role in developing predictive models of chemical reactivity \cite{nguyen_many_2014}. Using informative electronic (e.g., Fukui functions \cite{fukui_frontier_1997,ayers_variational_2000}) and steric (e.g., Sterimol \cite{verloop_sterimol_1983,unger_quantitative_2007}) descriptors can help model generalization and performance, especially in low data environments. Given suitable descriptors and holding other process parameters constant, complex properties have been described with linear or nearly-linear models, e.g., catalyst performance and enantioselectivity  \cite{oslob_steric_1997, milo_organic_2015, sigman_development_2016, zahrt_prediction_2019, reid_holistic_2019} (Figure~\ref{fig:reid_holistic_2019}). Descriptors tailored to a specific reaction class can be effective representations for predicting regioselectivity \cite{banerjee_machine_2018} and yield \cite{ahneman_predicting_2018} among other performance metrics, although they may not be broadly applicable across reaction and substrate classes. In principle, these descriptors could be calculated with greater universality than expert-selected ones already known to be relevant \cite{landrum_machine-learning_2004}. Similarly, selectivity in complex synthetic steps can be explained by expert-defined DFT calculations \cite{elkin_computational_2018} that could, in principle, be automated.

\begin{figure}[h]
  \centering
  \includegraphics[width=\twocolumnsize]{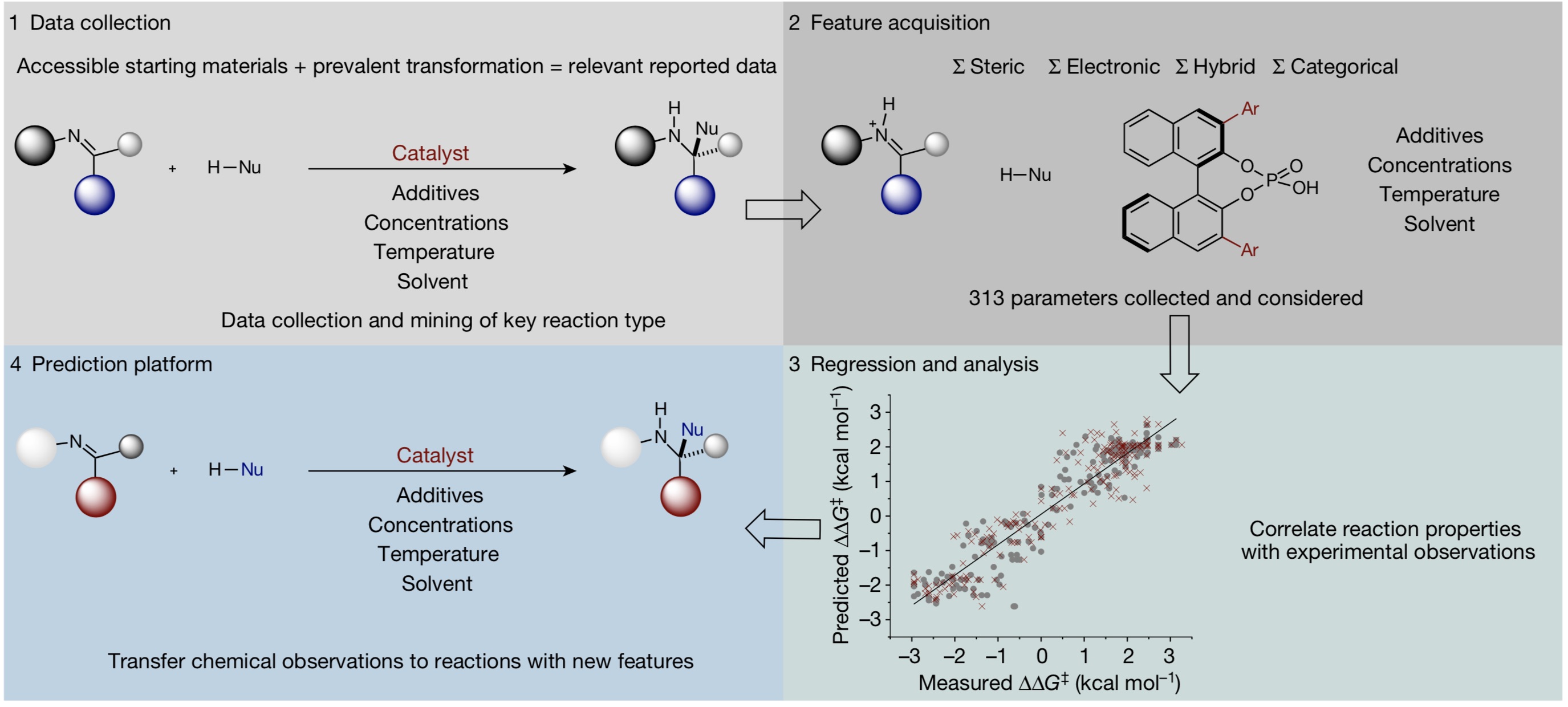}
  \caption{Discovery of catalysts for enantioselective catalysis using a surrogate model trained on experimental $\Delta \Delta G$ values to screen unseen reaction conditions. Figure reproduced from \citeauthor{reid_holistic_2019} \cite{reid_holistic_2019}.}
  \label{fig:reid_holistic_2019}
\end{figure}

If these models are truly describing the underlying patterns of chemical reactivity, they could be applied prospectively to the discovery of new synthetic methods. This is yet to be demonstrated. Time-split validations arguably demonstrate this generalization ability, however, a separate algorithm (a hypothesis generator) would be required to ``steer'' these models toward the combination of reactants mostly likely to result in new chemistry.

\subsubsection{Discovery of new chemical reactions from experimental screening} %
The discovery of new chemical reactions can widen synthetically-accessible chemical space and allow us to realize molecules that were previously difficult to access \cite{Blakemore2018,bostrom_expanding_2018}. The rise of combinatorial chemistry in the 1990s opened up new means of discovering new chemical reactions and functional physical matter through experimental screening \cite{brenner_encoded_1992,lehn_dynamic_1999}.  Low-volume liquid handling and rapid analysis by HPLC or ESI-MS or even fluorescent readouts \cite{shaughnessy_fluorescence-based_1999} have enabled  material-efficient reaction screening toward this end \cite{reetz_method_1999}. Microplate reaction screening has advanced to the point where it requires only nanomole quantities of material and achieves throughputs of thousands of reactions per hour \cite{Santanilla2015,lin_mapping_2018}; related technologies using continuous flow \cite{Perera2018} and electrospray ionization \cite{wleklinski_high_2018} can achieve similar throughputs and material consumption. 

These technologies have accelerated the rate at which candidate reactions (different substrates, conditions) can be tested, but still navigate a search space in a brute force manner.  High throughput experimentation \emph{can} be hypothesis-driven and used to investigate a narrower search space \cite{isbrandt_high_nodate, shevlin_nickel-catalyzed_2016} or be informed by mechanistic knowledge \cite{teders_diverse_2017} and functional diversity \cite{s.kutchukian_chemistry_2016, beeler_discovery_2007} (Figure~\ref{fig:kutchukian_chemistry_2016}), though this is less common in practice. Beyond improving the speed of experimentation and sensitivity of analysis, progress toward automated discovery of new chemical reactions has included developing new techniques for exploring the vast space of possible chemical reactions with fewer individual experiments: either by an active search (described later) or by pooling. Clever pooling strategies can allow for the simultaneous evaluation of multiple hypotheses through techniques like mass-encoded libraries \cite{geysen_isotope_1996}, DNA-templated synthesis \cite{kanan_reaction_2004}, and substrate combinations designed to enable straightforward deconvolution \cite{robbins_simple_2011, troshin_snap_2017} for ``accelerated serendipity'' \cite{mcnally_discovery_2011}. Multicomponent reactions represent a particularly large space and have historically been discovered either through serendipity or pooled/combinatorial screening \cite{weber_discovery_1999, domling_discovery_2000, ganem_strategies_2009}. \citeauthor{collins_contemporary_2014} review screening approaches to reaction discovery and development \cite{collins_contemporary_2014}.

\begin{figure}[h]
  \centering
  \includegraphics[width=\onecolumnsize]{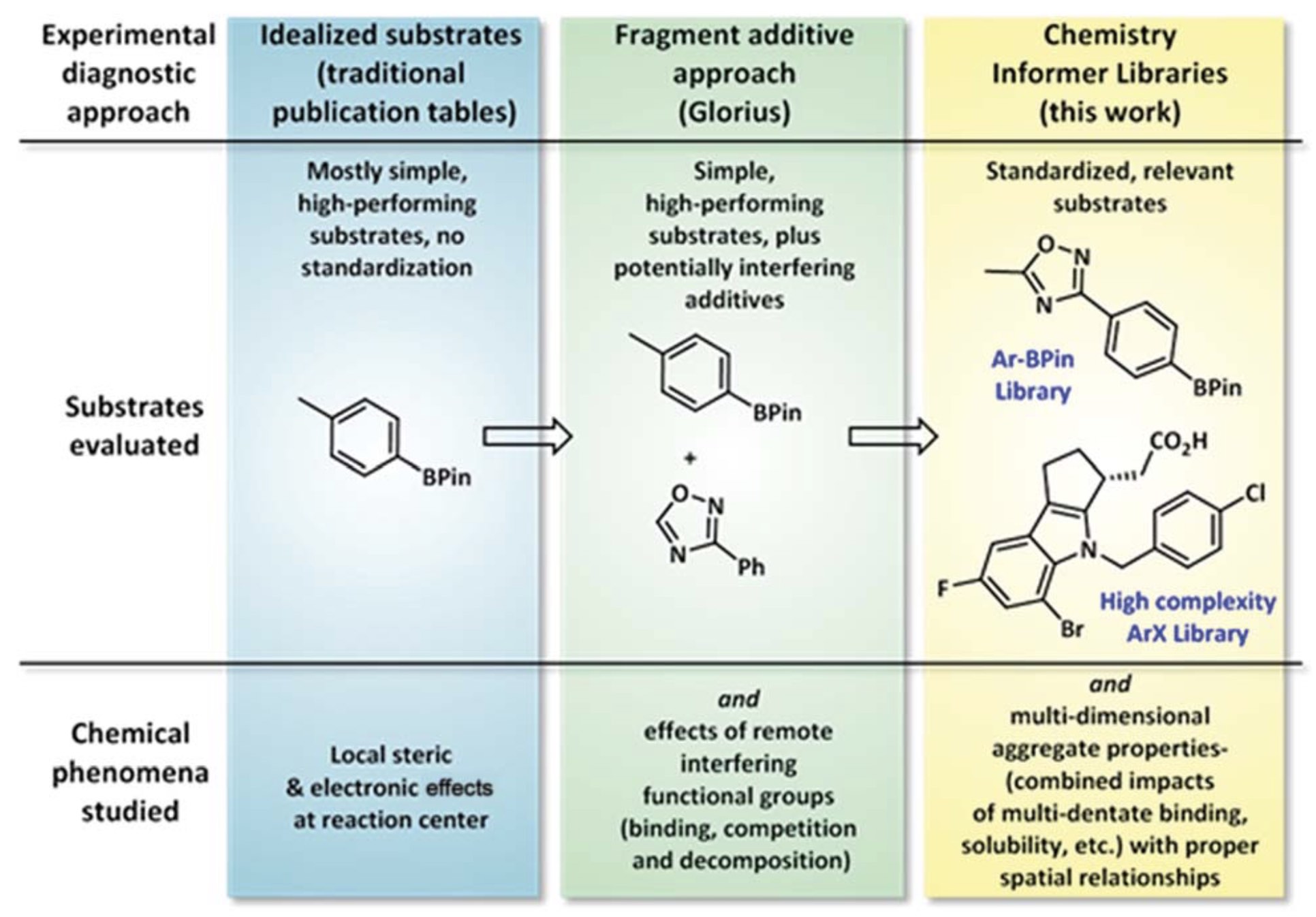}
  \caption{Chemistry informer approach to reaction screening to understand substrate compatibility, emphasizing the use of complex substrates to understand more complex chemical phenomena. Figure reproduced from \citeauthor{s.kutchukian_chemistry_2016} \cite{s.kutchukian_chemistry_2016}.}
  \label{fig:kutchukian_chemistry_2016}
\end{figure}

\subsection{\emph{Iterative} discovery of chemical processes}
\label{sec:cases:iterative_process}

\subsubsection{Discovery of optimal synthesis conditions}
Automatic discovery of optimal synthesis conditions is a  task where closed-loop experimentation is frequently applied. With a platform able to perform reactions under a wide range of operating conditions and automatically analyze and interpret the outcomes, one can use an optimization algorithm to guide a search within a pre-defined process parameter space. As with many other examples of automated discoveries, the search space is highly constrained by expert human operators. 
It is the case that standard numerical optimization routines are  sufficient to explore the narrow search space of interest when an expert is able to define a narrow range of conditions that will likely lead to promising results. 
The earliest automated platforms for organic reactions used batch reactors and computer-controlled valves or pumps to automatically add reagents according to computer-selected experiments \cite{deming_automated_1971, winicov_chemical_1978}.  Automated control of continuous process variables (e.g., residence time, temperature, reactant ratios) is simplified when using flow platforms that eliminate the need to physically replace or clean batch vessels. Due to the ease of sampling a crude product stream with an inline valve, they are frequently used to screen arrays of different process conditions to map out an experimental space and the corresponding parameter-performance relationship \cite{sugimoto_automated-flow_2009, koch_optimizing_2009, nieuwland_fast_2011, browne_continuous_2012}. Automated optimization of organic reactions in flow (Figure~\ref{fig:mateos_automated_2019}) has been extensively reviewed and is an excellent entry point for those interested in automated chemistry \cite{l.hartman_microchemical_2009, fabry_self-optimizing_2014, ley_organic_2015, houben_automatic_2015, mohamed_reaction_2016, reizman_feedback_2016, sans_towards_2016,mateos_automated_2019, horbaczewskyj_introduction_2019}.

\begin{figure}[h]
  \centering
  \includegraphics[width=\onecolumnsize]{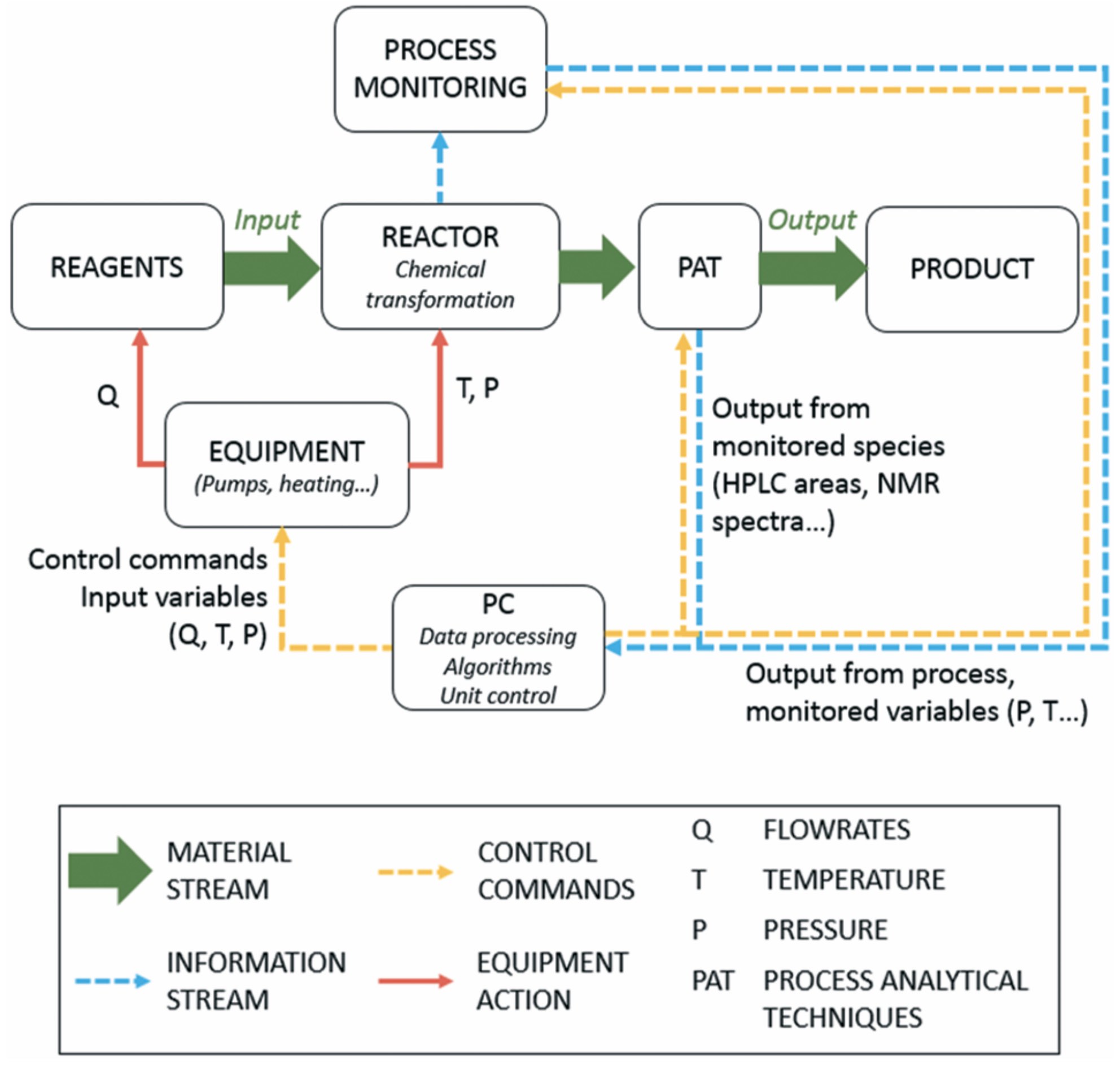}
  \caption{General platform schematic for the iterative optimization of synthetic processes in flow with respect to continuous variables (flowrates, temperature, pressure). Figure reproduced from \citeauthor{mateos_automated_2019} \cite{mateos_automated_2019}.}
  \label{fig:mateos_automated_2019}
\end{figure}

Optimization routines that have been widely employed include conjugate gradient \cite{fletcher_function_1964}, simplex \cite{nelder_simplex_1965}, genetic algorithms (GAs) \cite{holland_outline_1962}, Stable Noisy Optimization by Branch and Fit (SNOBFIT)  \cite{huyer_snobfitstable_2008}, adaptive response surface methods \cite{bezerra_response_2008}, Bayesian optimization approaches \cite{pelikan_boa:_1999}, and reinforcement learning \cite{sutton_reinforcement_2018}.
Optimizations over continuous variables generally use black box methods like genetic algorithms, Simplex, and SNOBFIT \cite{mcmullen_automated_2010,mcmullen_integrated_2010,parrott_self-optimizing_2011,skilton_real-time_2013,sans_self_2015,holmes_online_2016,poscharny_machine_2018,rubens_precise_2019}, gradient-based methods like steepest descent and conjugate gradient \cite{moore_automated_2012}, or explicit model-based methods like an adaptive response surface \cite{j.reizman_suzukimiyaura_2016,echtermeyer_self-optimisation_2017,m.baumgartner_optimum_2018}. In a recent study, \citeauthor{bedard_reconfigurable_2018} describe a reconfigurable flow platform that uses the SNOBFIT algorithm to optimize several  common organic transformations \cite{bedard_reconfigurable_2018}. Discrete process variables can be varied through the use of selector valves or liquid handlers \cite{hwang_segmented_2017, Perera2018} to also optimize, e.g., catalyst/ligand identities \cite{kreutz_evolution_2010,j.reizman_suzukimiyaura_2016,m.baumgartner_optimum_2018} and reaction solvent \cite{reizman_automated_2015}.  Despite their hype, ``modern'' machine learning approaches to reaction optimization \cite{zhou_optimizing_2017, Reker2018} have not demonstrated any clear advantages over previously-used statistical methods. One underexplored opportunity is to embed prior chemical knowledge into the model through pretraining; \citeauthor{zhou_optimizing_2017} do this not with chemical knowledge, but with knowledge about the geometry/roughness of the expected regression surface to improve the hill-climbing efficiency of a reinforcement learning optimization routine \cite{zhou_optimizing_2017}.

When the performance of a chemical process is measured by multiple objectives, it is important to understand their associated tradeoffs \cite{e.walker_tuning_2017}. Rather than combining them into a single scalar metric to optimize over \cite{houben_closed-loop_2015, fitzpatrick_novel_2016, hase_phoenics:_2018}, one can optimize for knowledge of the Pareto front--settings of process variables where one performance metric cannot be increased without decreasing another \cite{garg_multiobjective_1999, schweidtmann_machine_2018}. Multi-step reactions are particularly challenging to optimize, because the effects of changing one parameter can propagate through downstream process steps. They are typically broken up into individual synthetic steps to improve the tractability of the problem \cite{cortes-borda_autonomous_2018,wimmer_autonomous_2019} or optimized approximately through screening, rather than true closed-loop feedback \cite{sagmeister_laboratory_2019}.

Similar closed-loop optimizations have been demonstrated for  materials-focused applications. Different properties of interest necessitate different analytical endpoints, but the overall workflow is the same. Optimization goals have included the emission intensity of quantum dots \cite{krishnadasan_intelligent_2007}, the conversion and particle size resulting from a copolymerization \cite{houben_closed-loop_2015}, the identification of crystallization conditions for polyoxometalates \cite{Duros2017,Duros2019}, the production of Bose-Einstein condensates \cite{wigley_fast_2016}, and the realization of a metal-organic framework (MOF) with high surface area \cite{moosavi_capturing_2019}. The MOF synthesis optimization by \citeauthor{moosavi_capturing_2019} is particularly noteworthy in that prior data on syntheses of other MOFs were used to estimate the relative importance of  synthetic parameters to enable a maximally diverse initial design of experiments, jump-starting the  phase of iterative empirical optimization \cite{moosavi_capturing_2019}.

The challenge for these discoveries is often practical, not methodological. Experimental platforms must be able to analyze the relevant performance metrics and to control process variables across a search space that is broad enough to make computational assistance worthwhile. The ARES (autonomous research system) is an example of how complex instrumentation can enable optimizations of processes that are traditionally difficult to automate \cite{nikolaev_discovery_2014, nikolaev_autonomy_2016}. ARES can perform up to 100 carbon nanotube growth experiments via chemical vapor deposition (CVD) per day under different temperatures, pressures, and gas compositions with real-time monitoring of growth rates using Raman spectroscopy. After fitting a random forest model with 84 expert-defined experiments as prior knowledge, a genetic algorithm was successfully applied achieve a user-defined target growth rate through automated control of process conditions.

Iterative discovery of quantitative models of process performance (e.g., experimentation to estimate kinetic parameters) differs from optimization only in how experiments are selected. Instead of selecting experiments with the ultimate goal of maximizing yield or achieving optimal product properties, experiments can be selected to minimize uncertainty in regressed parameters  or discriminate between multiple hypothesized models \cite{mcmullen_rapid_2011}. The acquisition functions needed for these goals--to quantify how useful a proposed experiment would be--can be directly imported from  work in statistics on parameter estimation and model discrimination \cite{anderson-cook_response_2009}. There are still challenges for multi-step reactions, as deconvoluting the effects of kinetic parameters from individual steps may not be straightforward even when the rate laws are known \cite{reizman_automated_2012}.

\subsubsection{Discovery of new chemical reactions through an active search} 

There are far fewer examples of trying to discover of new chemical reactions through  \emph{active} searches than through noniterative screening strategies. \citeauthor{amara_automated_2015} describe one example of discovering new reaction pathways in a catalytic reactor system by reformulating the problem as a reaction optimization \cite{amara_automated_2015}. Using a modified Simplex algorithm, they were able to optimize the yield of then-uncharacterized side products; mechanistic pathways were proposed by experts based on evaluation of the conditions leading to different product distributions.

\citeauthor{granda_controlling_2018} instead treat reaction discovery as a natural consequence of building a quantitative model of chemical reactivity \cite{granda_controlling_2018} (Figure~\ref{fig:granda_controlling_2018}). Specifically, they describe a platform for evaluating the reactivity of two- and three-component reactions among a set of 18 hand-picked building block molecules (969 possible experiments) using two empirical models: one makes a boolean prediction of whether a reaction has taken place based on NMR, MS, and ATIR data before/after mixing; the second makes a boolean prediction of whether a given combination of substrates is likely to be reactive, using a one-hot representation for chemical species. The physical apparatus is reminiscent of MEDLEY--an automated reaction system employing computer-controlled pumps connected to a round-bottom flask \cite{orita_automated_2000}. However, the goal of reaction discovery and training this binary classifier are misaligned: the algorithmic exploration of the search space of possible reactions does not direct experiments toward those likely to lead to novel reactions; the reactions claimed to have been discovered were identified only through (error-prone \cite{sader_reinvestigation_2019}) manual analysis of product mixtures. Moreover, all 969 reaction combinations could have been performed in a miniaturized well-plate format while using less time and materials (cf. item \ref{item:brute_force_search} of section \ref{sec:assessing_autonomy}) and the one-hot encoding of substrates precludes prediction of reactivity for unseen substrates. An earlier version of this platform was used by the same group to explore the 64 possible pathways defined by a three-step synthesis with one of four reagents added at each step \cite{dragone_autonomous_2017}. The ``most reactive'' pathway was found through a step-wise greedy search by identifying the reagent whose addition led to the largest change in the ATIR spectrum of the product mixture. While this too did not explicitly bias experiments towards new reactions, the concept of selecting experiments in a non-brute force manner for reaction discovery is worth further investment. 

\begin{figure}[h]
  \centering
  \includegraphics[width=\onecolumnsize]{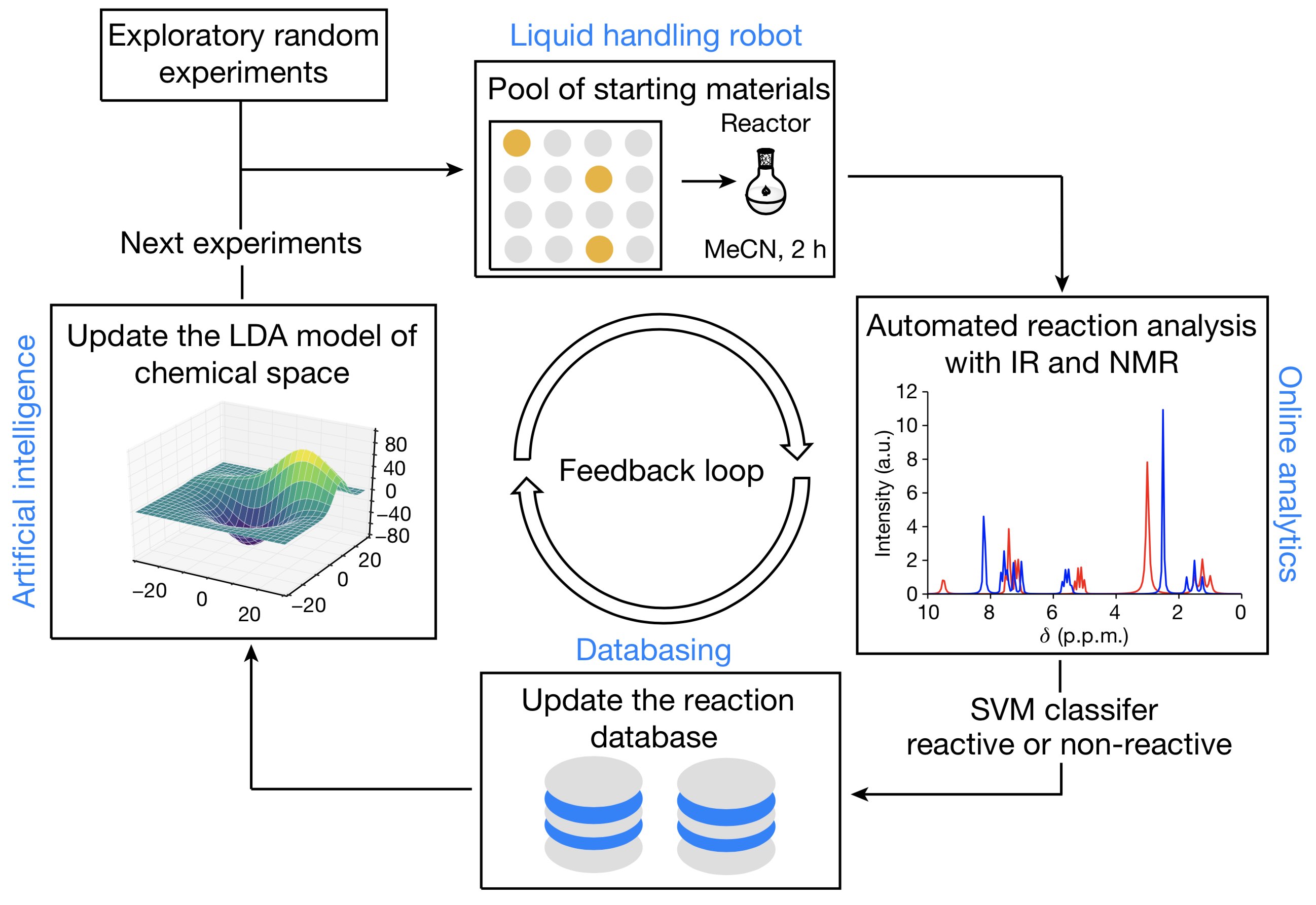}
  \caption{Workflow for iteratively training a binary classifier of whether a reaction mixture is reactive, using experimental validation and feedback. Figure reproduced from \citeauthor{granda_controlling_2018} \cite{granda_controlling_2018}.}
  \label{fig:granda_controlling_2018}
\end{figure}

\subsection{\emph{Noniterative} discovery of structure-property models}
\label{sec:cases:noniterative_qsar}

Models capable of relating the structural and compositional features of a molecule or material to its properties are of substantial utility in discovery. These relationships are often learned directly from data, whether via standard multivariate regression or machine learning algorithms.  
To the extent that they are interpretable, they can yield insight into how the fundamental features of a chemical entity or system influence its properties or performance, thus informing design. 
Quantitative structure-activity/property relationships (QSARs/QSPRs) can act as our belief about a performance landscape (cf. ``belief'' in Figure~\ref{fig:our_iterative_process_search}) for the sake of a specific discovery task like the discovery of new physical matter. While there is only a weak distinction between developing a QSAR/QSPR model for its own sake or for the purposes of exploring a design space, this section will focus on studies where the primary discovery is of the model itself. General considerations and trends in QSAR/QSPR are discussed in refs.~\citenum{dudek_computational_2006}, \citenum{Mitchell2014}, and \citenum{cherkasov_qsar_2014}.

\subsubsection{Discovery of important molecular features} 
Given a QSAR/QSPR model, one can investigate how the model perceives different structural attributes to reveal which are most informative of the prediction task. Substructure filters are commonly employed to process screening hits \cite{pearce_empirical_2006,baell_new_2010, baell_seven_2018} and flag reactive or toxic functional groups \cite{sanderson_computer_1991, kazius_derivation_2005, sushko_toxalerts:_2012,  basant_predicting_2015,metivier_discovering_2015,li_deepchemstable:_2019}. This is a standard problem of interpretability that has received significant attention in the machine learning community \cite{doshi-velez_towards_2017}. When the form of the desired interpretation is restricted to molecular substructures, then standard approaches for variable and feature selection can be applied when using a representation based on the presence/absence of certain substructures \cite{tibshirani_regression_1996, guyon_introduction_2003}. \citeauthor{polishchuk_interpretation_2017} provides a recent review of interpretability for QSAR/QSPR models, including this category \cite{polishchuk_interpretation_2017}.

 An early attempt to correlate predicted function directly with structural attributes was PROGOL \cite{king_structure-activity_1996}, an inductive logic programming algorithm. In its original demonstration, PROGOL identified a set of five criteria for determining whether a compound is likely to be mutagenic based on the presence of hypothesized toxicophores defined by connectivity and partial charge values; subsequent studies pursued similar explanations for carcinogenicity \cite{king_prediction_1996} and ACE inhibition activity \cite{finn_pharmacophore_1998}, among others (Figure~\ref{fig:king_prediction_1996}).

\begin{figure}[h]
  \centering
  \includegraphics[width=4cm]{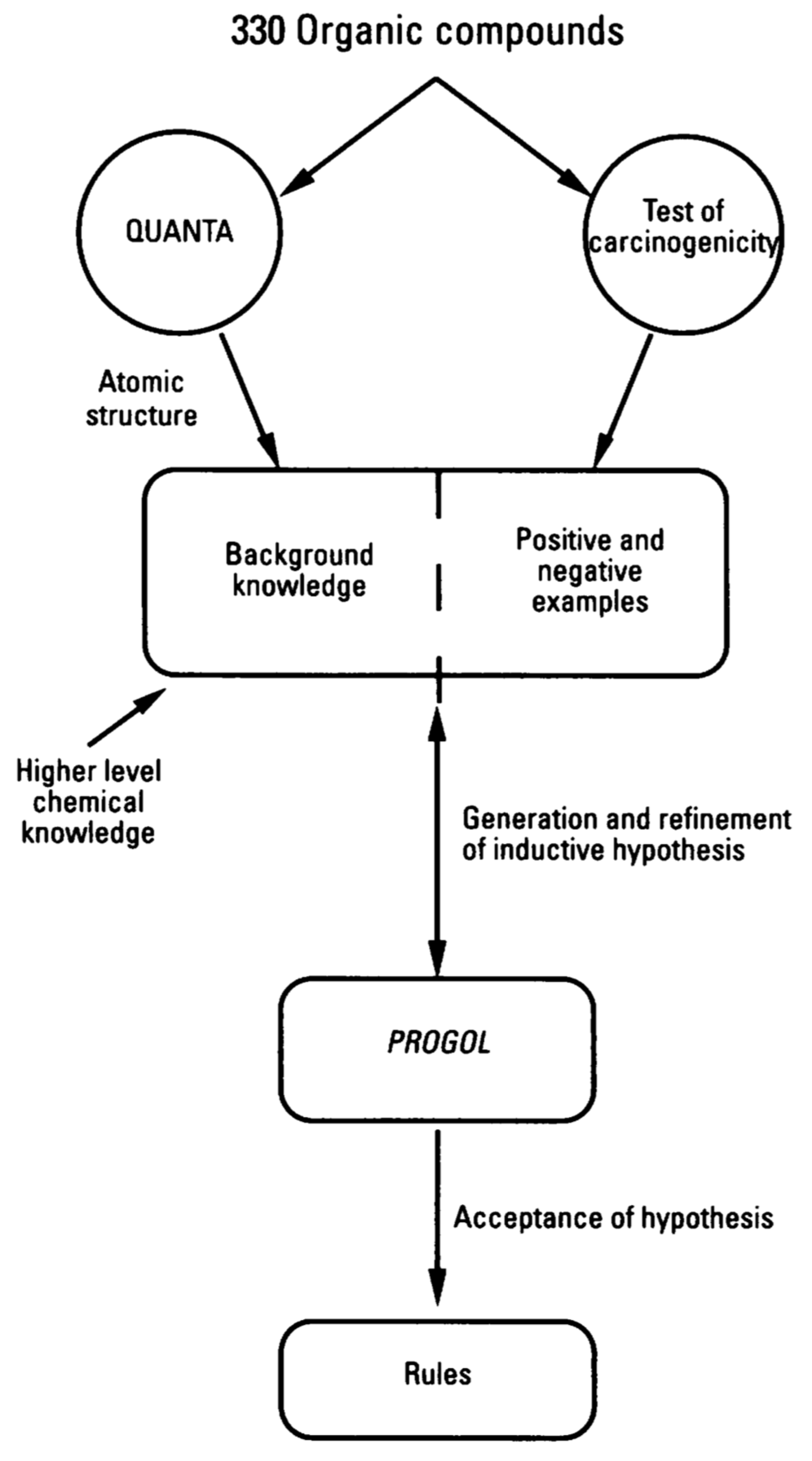}\hspace{0.5cm} \includegraphics[width=9cm]{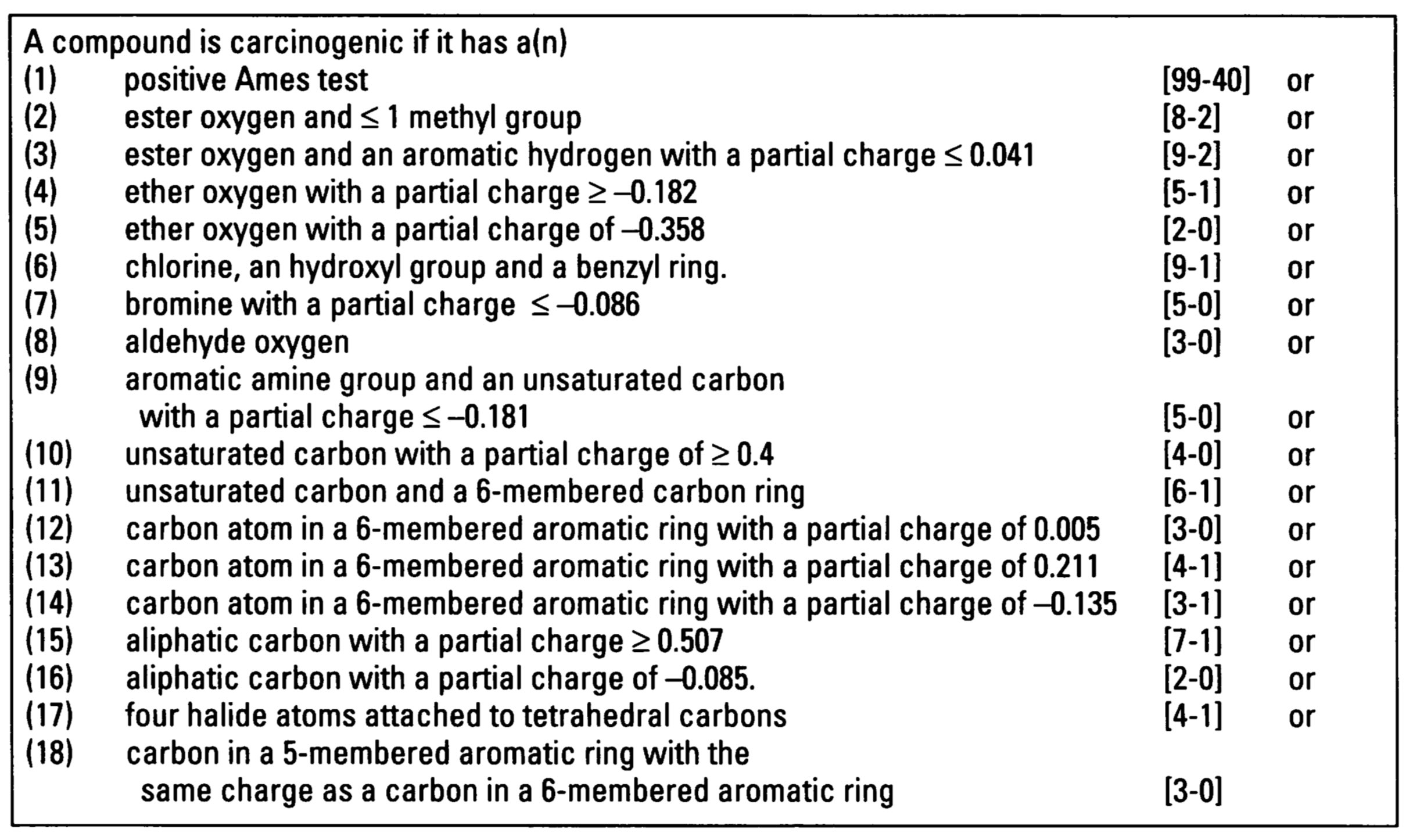}
  \caption{Workflow for the application of PROGOL \cite{king_structure-activity_1996} to the induction of structural alerts for carcinogenicity and the resulting rules. [$x$-$y$] indicates that the rule correctly applied $x$ times and incorrectly applied $y$ times in the dataset of 330 organic compounds. Figure reproduced from \citeauthor{king_prediction_1996} \cite{king_prediction_1996}.}
  \label{fig:king_prediction_1996}
\end{figure}

One approach to interpretability is to rely on few-parameter regressions with interpretable descriptors \cite{hansch_correlation_1962,free_mathematical_1964}, which can provide explanations as meaningful as the descriptors themselves. Decision trees provide a natural mode of assessing descriptor importance, though ensembling methods (e.g.,  Random Forest models) can obfuscate analysis  \cite{svetnik_random_2003,kuzmin_interpretation_2011}. More general techniques exist to extract symbolic rules from trained machine learning models that are relatively agnostic to the type of model used \cite{towell_extracting_1993,barakat_rule_2010, preuer_interpretable_2019}; ref. \citenum{raccuglia_machine-learning-assisted_2016} provides an example of a decision tree extracted from an SVM model trained to predict the success of an inorganic synthesis procedure. There are numerous other examples of QSAR/QSPR studies that estimate descriptor importance in an attempt to rationalize predictions \cite{byvatov_svm-based_2004, eklund_choosing_2014,  coley_convolutional_2017, lee_ligand_2019, li_making_2019, polishchuk_interpretation_2017}. Other approaches instead aim to identify the training examples most relevant to a given test example \cite{hansen_visual_2011, koh_understanding_2017}.

Visualizing explanations can be more intuitive than looking at quantitative feature importance metrics. One popular approach is to approximate a model by a fragment-contribution approach by looking at how the predicted property changes when part of the input molecule is masked \cite{riniker_similarity_2013, polishchuk_universal_2013, polishchuk_structural_2016}. If the value decreases when masking a certain substructure, that substructure is assumed to positively contribute to the property. This per-atom or per-substructure importance metric is usually an oversimplification of what is being learned, though sometimes it is exactly what is being learned \cite{schutt_quantum-chemical_2017}. The accuracy of machine learning models is usually at least partially attributable to the nonlinearities between the input featurization and output property. 

\subsubsection{Discovery of models for spectral analysis} %
A natural application of data science techniques is to the analysis of spectral data for computer-aided structural elucidation (CASE). The underlying function that maps a molecule or material to the results of an assay is no different than a standard structure-property relationship, except the property might be high dimensional. CASE will become increasingly important for structure confirmation and quantitation as autonomous systems start to explore new areas of chemical and reactivity space.

The DENDRAL program is an early example of a program designed for structural elucidation of organic structures from mass spectrometry (MS) data \cite{buchanan_heuristic_1969,feigenbaum_generality_1971}. It crossreferences the mass loss between peaks with a list of known fragments to identify the likely substituents of the original molecule, enumerates possible molecular structures, predicts the MS spectra of those candidate structures, and makes its final proposal based on consistency with the observed spectrum. DENDRAL proved useful in its ability to perform many rapid calculations (spectral simulations and matching), but still required expert heuristics to help explore the vast space of possible molecular structures, including a ``badlist'' to prune unrealistic structures.  \citeauthor{hufsky_mining_2017} provide a recent review of computational analysis of MS fragmentation patterns \cite{hufsky_mining_2017}. Computational MS analysis continues to be the application of supervised learning approaches \cite{wei_predicting_2018} and has seen renewed interest in the context of metabolomics \cite{allen_competitive_2015,blazenovic_software_2018,samaraweera_evaluation_2018,ludwig_bayesian_2018}.  

Unsurprisingly, other types of analytical data are also commonly evaluated using computational or machine learning models, including UV circular dichroism to elucidate protein secondary structure \cite{bohm_quantitative_1992}. The reverse problem of spectral prediction is also popular and includes techniques to predict NMR shifts \cite{aires-de-sousa_prediction_2002,binev_prediction_2007}, IR spectra  \cite{gastegger_machine_2017}, and protein fluoresence \cite{nantasenamat_prediction_2007}. Materials-focused studies have looked at predicting the optical properties of metal oxides \cite{s.stein_machine_2019} and analyzing microstructure from SEM data \cite{ling_building_2017}, among others. Tables 1 and 2 of ref. \citenum{sumpter_design_1996} summarize many early examples of applying neural networks to MS, NMR, IR, NIR, UV, and fluoresence spectra leading up to the mid-1990s. A more recent overview of learning structure-spectrum relationships and CASE can be found in ref.~\citenum{engel_structure-spectrum_2018}.

\subsubsection{Discovery of potential energy surfaces and functionals}
There is tremendous interest in using machine learning techniques to build surrogate models for computationally-expensive \emph{ab initio} calculations.  Models can replace either the entire energy calculation \cite{behler_generalized_2007, montavon_machine_2013, hansen_assessment_2013, schmitz_machine_2019} or specific parameterized components (e.g., functionals or correlation energies) \cite{snyder_finding_2011, behler_first_2017,  welborn_transferability_2018, cheng_universal_2019}. A prominent example from Roitberg, Isayev, and coworkers is ANAKIN-ME or ANI (accurate neural network engine for molecular energies); ANI is a neural network surrogate model of an energy potential trained on roughly 60,000 DFT calculations \cite{smith_ani-1:_2017} and its second generation, ANI-1ccx, is further refined on a smaller set of CCSD(T)/CBS calculations \cite{s_smith_outsmarting_nodate} (Figure~\ref{fig:s_smith_outsmarting_nodate}).  Active learning strategies can be used to strategically acquire costly training data when training such models \cite{podryabinkin_active_2017, smith_less_2018, s_smith_outsmarting_nodate}. The accurate prediction of electronic properties is directly useful for the discovery of organic electronic materials and is a central focus of the Harvard Clean Energy Project \cite{hachmann_harvard_2011}.  %

\begin{figure}[h]
  \centering
  \includegraphics[width=\onecolumnsize]{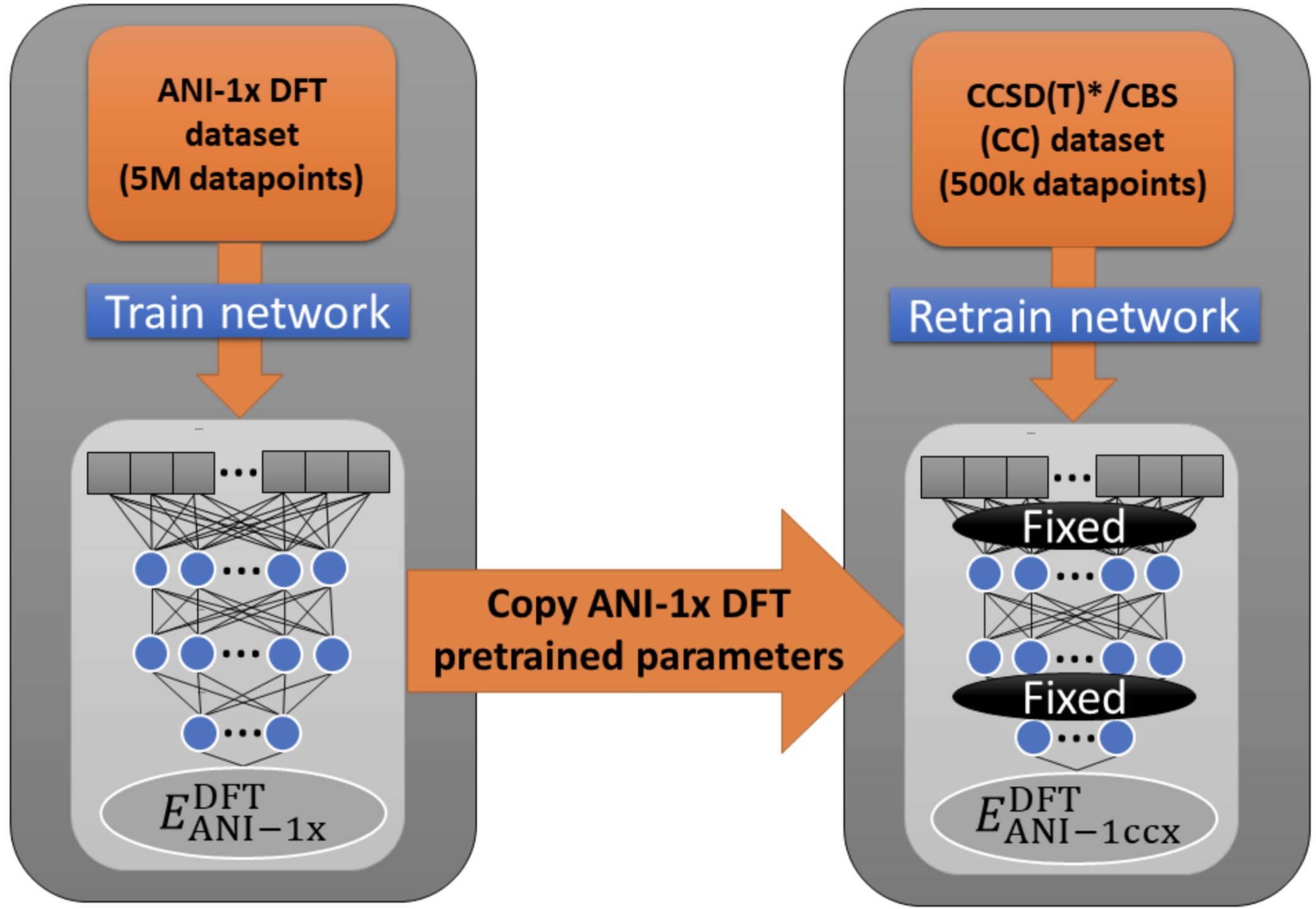}
  \caption{Workflow for refining a surrogate model of electronic structure calculations, originally trained on $\omega$B97x/6-31G* data, on higher-quality CCSD(T)*/CBS data. Figure reproduced from \citeauthor{s_smith_outsmarting_nodate} \cite{s_smith_outsmarting_nodate}.}
  \label{fig:s_smith_outsmarting_nodate}
\end{figure}

The desire to create computationally-inexpensive surrogate models has underpinned the development of classical force field models for molecular dynamics (MD) simulations \cite{mayo_dreiding:_1990, rappe_uff_1992, wang_development_2004}. Perhaps unsurprisingly, machine learning models can serve as drop-in replacements for heuristic force fields if so trained \cite{bartok_gaussian_2010, zhang_deep_2018, deringer_realistic_2018} and can assist in coarse-graining for larger scale simulations \cite{wang_coarse-graining_2018,wang_machine_2019} or as a post-processing step to analyze simulation results \cite{f.reinhart_machine_2017,mardt_vampnets_2018}. Structure-aided drug design relies on similar parametric functions for predicting protein-ligand binding. There are many molecular docking programs that propose and score different poses describing ligand interactions with protein targets \cite{warren_critical_2006, pagadala_software_2017, su_comparative_2019}. Scoring functions--meant to provide quantitative measures that correlate with binding affinity--are ideal applications of machine learning techniques. Nonlinear statistical models can help bridge the divide between our pseudo-first principles models of the underlying chemical interactions and the actual behavior we observe experimentally \cite{ballester_machine_2010, ain_machine-learning_2015, pereira_boosting_2016, bjerrum_machine_2016, wojcikowski_performance_2017, jimenez_kdeep:_2018, stepniewska-dziubinska_development_2018, su_comparative_2019}.

\subsubsection{Discovery of models for phase behavior} 
QSAR/QSPR models that describe a molecule or material's phase behavior can aid computational design by predicting whether a proposed compound is physically realizable in its desired form. For hypothesized metallic alloys, for example, one can predict crystal structures \cite{curtarolo_predicting_2003,fischer_predicting_2006, ulissi_automated_2016, ziletti_insightful_2018} and phase behavior \cite{levy_uncovering_2010, hautier_finding_2010, meredig_combinatorial_2014, gautier_prediction_2015, artrith_constructing_2018, nguyen_hybrid_2018}. For organic molecules, one can similarly predict whether compounds are likely to crystallize easily \cite{wicker_will_2015} and their preferred processing-dependent polymorph \cite{hiszpanski_data_2018}. Machine learning models can also reduce the number of evaluations required, e.g., for finding minimum energy configurations \cite{ulissi_automated_2016}.

\subsection{\emph{Noniterative} discovery of new physical matter}
\label{sec:cases:noniterative_phys}

The noniterative discovery of new physical matter is a common application of computational learning techniques or automated experimental platforms. This category encompasses experimentation strategies in which search spaces are predefined and explored exhaustively and virtual screening with or without the use of a surrogate model to approximate a structure-function landscape (right half of Figure~\ref{fig:taxonomy_disc_phys_matt}).

A quintessential paradigm in this category is the use of a large dataset from experiments or simulations to train a QSAR/QSPR model, often using some form of machine learning for nonlinear regression, which is then used to screen a large number of candidate compounds or materials. A handful of candidates may be selected for synthesis and validation of the prediction, but the results of that validation are not used to revise the model. This approach essentially constructs a fixed ``map'' with which to explore the search space and identify promising candidates. It leaves little room for serendipity, as compounds that are not predicted to be useful--even if accounting for uncertainty--are generally not tested, unless the algorithm is explicitly biased toward random exploration.

\begin{figure}[h]
  \centering
  \includegraphics[width=\twocolumnsize]{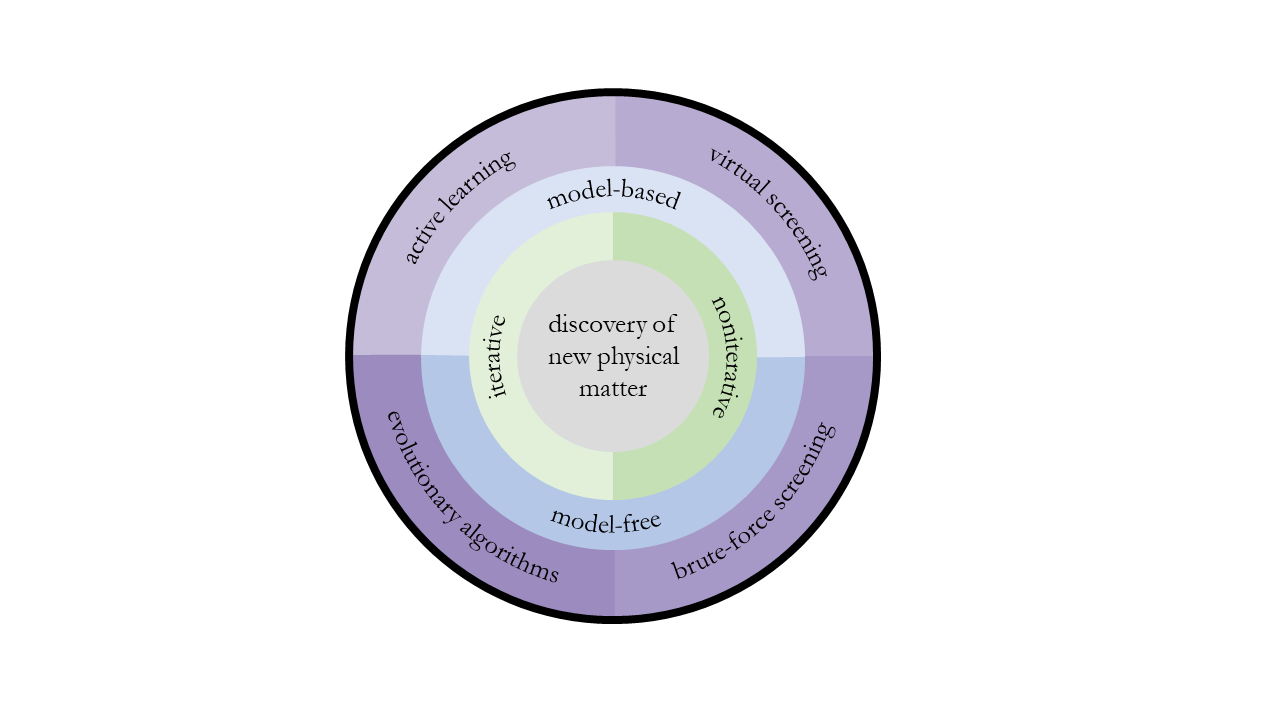}
  \caption{Taxonomy of strategies for the discovery of new physical matter.}
  \label{fig:taxonomy_disc_phys_matt}
\end{figure}

\subsubsection{Discovery through brute-force experimentation}
In several studies, the part of the discovery process that is automated is not the hypothesis (model building or selection of compounds to test) but the experiment (initial data generation or validation). Experimental automation  addresses the practical challenge of validation but not the methodological challenge of how to guide the scientific process or constrain the search space. In general, brute-force experimentation is a productive  discovery strategy only when the experimentation is high-throughput in nature. High-throughput experimentation platforms are capable of searching broad design spaces, which makes serendipitous discoveries more likely and places less emphasis on the experiment-selection faculties of the researchers. Note, however, that \emph{manual} constraint of the design space remains a critical aspect of the process. An ideal high-throughput platform is high-throughput from end to end (synthesis to analysis), but many of the case studies described herein emphasize the development of tools that accelerate just one of these stages.

Among the more interesting developments in HTE for drug discovery are entirely novel methodologies that are uniquely suited to rapid data generation. Despite the advances in achieving greater throughput with traditional HTE efforts \cite{mayr_novel_2009, janzen_screening_2014}, the space that can be feasibly screened using single-compound-per-well synthesis approaches is often too small to provide many promising bioactive leads. DNA-encoded libraries (DELs), a concept introduced by \citeauthor{brenner_encoded_1992} \cite{brenner_encoded_1992}, enable synthesis of compounds for screening at rates of hundreds of compounds per well  \cite{scheuermann_dna-encoded_2006, mannocci_20_2011}. An adaptation of the split-and-pool synthesis strategy \cite{Thompson1996}, many modern DEL case studies report theoretical library complexities of hundreds of millions \cite{Clark2009,Litovchick2015,Ding2015} or even billions \cite{Harris2016,Kollmann2014,Deng2012} of compounds, exceeding the size of the search space  by several orders of magnitude over traditional HTE approaches \cite{GoodnowJr2017}. In light of these impressive synthesis rates, it should be noted that analysis and (if necessary) purification can be rate-limiting.  Recent successes of this strategy include the identification of a series of receptor interacting protein 1 (RIP1) kinase inhibitors \cite{Harris2016}, which are implicated in multiple inflammatory diseases, and an inhibitor of soluble epoxide hydrolase (sEH) with relevance to several disease areas \cite{Belyanskaya2017}. DNA-encoded chemistry is reviewed in ref.~\citenum{GoodnowJr2017}.

Another strategy that has proved useful in this area is diversity-oriented synthesis (DOS) which aims to generate structurally (and thereby functionally) diverse collections of small molecules \cite{Tan2005}. These strategies may involve reacting a starting material with diverse arrays of reagents in series, or coupling an array of starting materials to one another across strategic functional groups \cite{Galloway2010}; multicomponent reactions are particularly useful for this application \cite{Biggs-Houck2010}. \citeauthor{Tan2005} \cite{Tan2005}, \citeauthor{Spandl2008} \cite{Spandl2008}, \citeauthor{Galloway2010} \cite{Galloway2010}, and \citeauthor{garcia2016scaffold} \cite{garcia2016scaffold} review DOS strategies in detail and describe successful applications for the discovery of lead drug compounds and biological probes.

Other novel approaches to making HTE for drug discovery more efficient focus not on synthesizing larger and/or more diverse libraries, but on screening them more efficiently \cite{gesmundo_nanoscale_2018,Krutzik2006,Swann2016,Inglese2006,Guo2012}. Development of information-rich, efficient assays is a complex challenge. If the disease target has already been identified and it is possible to prepare the target such that it can be sufficiently isolated, stabilized, and accurately dispensed, then {\it{in vitro}} biochemical assays, which may involve the assessment of target-ligand binding affinity or augmentation of enzymatic activity, are useful {\cite{Walters2003,Schenone2013,Frearson2009}}. These assays can be easily miniaturized and serve as the workhorse of many screening campaigns. These target-based assays tend to be efficient, but they assess activity against a single target in isolation and ignore the complexities of human physiology and polypharmacology. Cell-based assays can do a better job capturing activity (because, for example, relevant cofactors are present) while also providing a measure of toxicity and other off-target effects. Despite the added complexity of automatically maintaining and dispensing cell populations, cell-based assays have been adequately automated and miniaturized for compatibility with the highest-density plates {\cite{Brandish2006}}. A variety of easily-automated well measurement tools have been developed for compatibility with cell-based assays, including fluorescent detection; the automation and miniaturization of this type of assay for compatibility with HTS has been well-reviewed by {\citeauthor{An2010}} {\cite{An2010}}. Recently, there has been an increased interest in phenotypic screening \cite{moffat_phenotypic_2014} along with an increased reliance on computational tools for high-throughput analysis \cite{carpenter_cellprofiler:_2006, jones_scoring_2009}. 

Many high-throughput synthesis and analysis tools have been developed to facilitate experimentation and discovery in materials science. Combinatorial synthesis methods that yield a single sample containing continuous composition gradients \cite{Green2013} are one notable example. In the seminal demonstration of this technique, \citeauthor{Xiang1995} describe a parallelized synthesis method for superconducting copper oxide thin films that varies composition, stoichiometry, and deposition sequence to identify promising compositions \cite{Xiang1995}. As \citeauthor{Senkov2015} point out, this type of experimentation can present unique obstacles to miniaturization in subdomains such as metal alloy design, where experiments of a certain scale are required to observe emergent macro- and mesoscale properties \cite{Senkov2015}. Since this early demonstration, combinatorial and other HTE methods have been developed in a wide array of materials subfields: to screen solid-state catalyst libraries \cite{Senkan1998,Senkan1999} as well as to discover cobalt-based MOFs \cite{wollmann_high-throughput_2011}, photosensitizers for catalyzing photoinduced water reduction \cite{Goldsmith2005}, mixed metal oxide catalysts \cite{Bergh2003}, adhesive coatings for automotive applications \cite{Potyrailo2003}, polymers for gene delivery \cite{Akinc2003}, ternary alloys that have high glass forming ability \cite{Tsai2016}, and others.

Development of high-throughput analytics is critical to avoid analytical bottlenecks; one method highly amenable to parallelization is infrared imaging \cite{Senkan1998,Holzwarth1998, Snively2001}, which has been used to screen arrays of heterogeneous catalysts \cite{Caruthers2003}. \citeauthor{potyrailo2005role} emphasize the diverse properties materials exhibit and the need for a correspondingly diverse array of characterization tools  \cite{potyrailo2005role}. As an alternative to developing high-throughput characterization techniques, \citeauthor{sun_accelerated_2019} recently reported the rapid exploration of a 96-member library of perovskite-inspired compositions (Figure~\ref{fig:sun_accelerated_2019}) in which they accelerated screening in part by replacing rate-limiting analytics with a cheaper analysis and a machine learning model \cite{sun_accelerated_2019}. Their study included the first reporting of four lead-free compositions in thin-film form. Reviews pertinent to high-throughput experimental screening for materials discovery are provided by \citeauthor{Meier2004} \cite{Meier2004}, \citeauthor{Zhao2006} \cite{Zhao2006}, \citeauthor{Rajan2008} \cite{Rajan2008}, \citeauthor{hook_high_2010} \cite{hook_high_2010}, \citeauthor{potyrailo_combinatorial_2011} \cite{potyrailo_combinatorial_2011}, and \citeauthor{Green2013} \cite{Green2013}.

\begin{figure}[h]
  \centering
  \includegraphics[width=14cm]{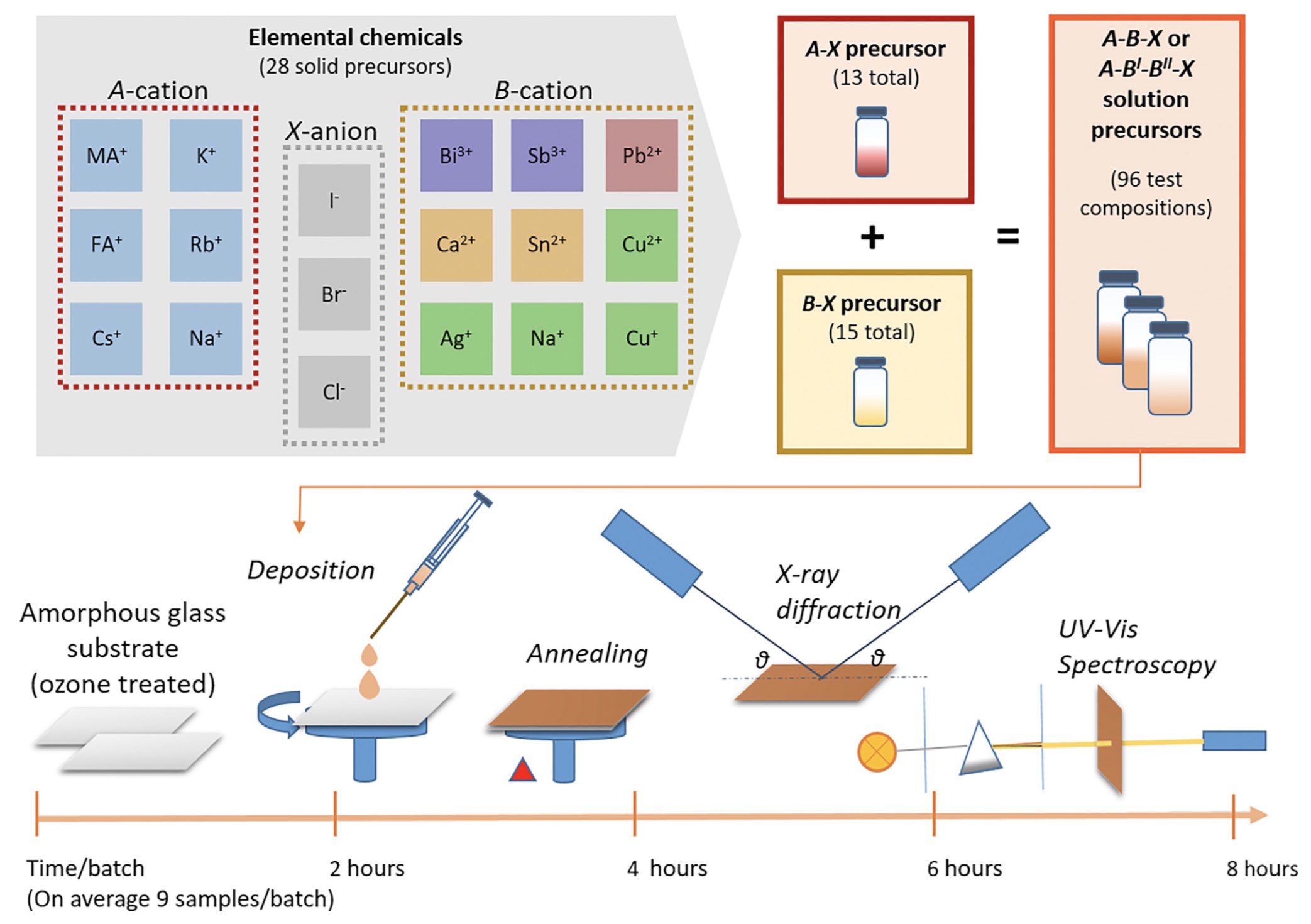}
  \caption{Workflow for the (relatively) rapid screening of a combinatorial space of perovskite-like compositions. Figure reproduced from  \citeauthor{sun_accelerated_2019} \cite{sun_accelerated_2019}.}
  \label{fig:sun_accelerated_2019}
\end{figure}

\subsubsection{Discovery through computational screening}
There are a number of noniterative approaches to materials discovery \cite{norskov_towards_2009} that rely on computational workflows like virtual screening or high-throughput simulation techniques \cite{Setyawan2010,pyzer-knapp_what_2015,jain_commentary:_2013}; \citeauthor{Ong2013} developed a Python library (pymatgen) specifically designed to facilitate these workflows \cite{Ong2013}. \citeauthor{curtarolo_high-throughput_2013}, \citeauthor{pyzer-knapp_what_2015}, and \citeauthor{himanen_data-driven_2019} provide recent reviews on high throughput virtual screening for materials discovery, specifically describing the process of generating large databases and identifying meaningful trends within them \cite{curtarolo_high-throughput_2013, pyzer-knapp_what_2015, himanen_data-driven_2019}.

The predominant use of machine learning in computational materials discovery has been to fit surrogate models to existing (often, experimental) data and screen a large design space \cite{bhadeshia_performance_2009, fjell_identification_2009, le_quantitative_2012, rajan_materials_2015, zhang_data_2019, gu_machine_2019, freeze_search_2019}.  To the extent that performance can be correlated to structure, these models can reveal opportunities for the design of new catalysts/ligands for organic synthesis \cite{oslob_steric_1997, sigman_development_2016, meyer_machine_2018, zahrt_prediction_2019, reid_holistic_2019} (Figure~\ref{fig:reid_holistic_2019}), metallic catalysts \cite{li_high-throughput_2017, jinnouchi_predicting_2017}, Heusler compounds \cite{oliynyk_high-throughput_2016}, metal organic frameworks (MOFs) \cite{rosen_identifying_2019}, hybrid organic-inorganic perovskites \cite{lu_accelerated_2018}, superhard materials \cite{mansouri_tehrani_machine_2018}, thermal materials \cite{zhang_machine_2019}, organic electronic materials \cite{Hachmann2014,pyzerknapp_learning_2015, gomez-bombarelli_design_2016, sun_accelerating_2018, lu_rapid_nodate}, polymers for electronic applications \cite{zeng_graph_2018, wilbraham_mapping_2019}, porous crystalline materials for gas storage \cite{Colon2014, pulido_functional_2017}, and reductive additives for battery electrolyte formulations \cite{Halls2010}. Computational models have also been used to determine when calculations are likely to fail \cite{duan_learning_2019} and to identify associations between materials and specific property keywords through text mining \cite{tshitoyan_unsupervised_2019}.

A few trends are apparent in the experimental validation of these frameworks. First, the confidence of computational predictions is intimately coupled to the quality and applicability of the model. Second, experimental validation is often preceded by extensive manual filtering of the computationally-prioritized compounds to take into account factors such as synthesizability, laboratory capabilities, and (human-)perceived suitability for the discovery objective \cite{nagasawa_computer-aided_2018,gomez-bombarelli_design_2016, Wilmer2012}. As with organic molecules, there can be a misalignment between the compounds one would like to test (that are computationally predicted to achieve a desired function) and what can be realized (synthesized) experimentally. In one example, as a conservative filter for synthesizability, \citeauthor{sumita_hunting_2018} require that proposed molecules have at least one known synthetic route reported in SciFinder \cite{sumita_hunting_2018}. Third, there are often discrepancies between the predictions made by a surrogate model and the values determined experimentally, and in some cases the discrepancies are large enough to have a substantial bearing on the desired performance \cite{nagasawa_computer-aided_2018}. These latter two trends imply that the pertinent features of promising materials are often not fully captured by the algorithms developed to date. Thus, experimental validation is acutely relevant in this area.

Computational screening, with or without experimental validation, is also a common strategy for identifying promising therapeutic candidates. Many reviews on the use of virtual screening in drug discovery exist, including  \citeauthor{schneider_virtual_2010} \cite{schneider_virtual_2010}, \citeauthor{sliwoski2014computational} \cite{sliwoski2014computational}, \citeauthor{macalino_role_2015} \cite{macalino_role_2015}, \citeauthor{Lavecchia2015} \cite{Lavecchia2015}, \citeauthor{Wingert2018} \cite{Wingert2018}, \citeauthor{zhang_machine_2017} \cite{zhang_machine_2017}, and \citeauthor{Panteleev2018} \cite{Panteleev2018}; these emphasize the use of machine learning methods to generate the surrogate QSAR/QSPR models that guide the VS process. \citeauthor{Walters2018} recently reviewed strategies for library enumeration in the drug discovery space \cite{Walters2018}. These range from applying known reaction transformations to available molecules in order to define make-on-demand libraries  \cite{chevillard_scubidoo:_2015,humbeck_chipmunk:_2018,vinkers_synopsis:_2003,Hartenfeller2012, nicolaou_proximal_2016, van_hilten_virtual_2019, lyu_ultra-large_2019} to generative strategies, which are discussed later in the context of \emph{de novo} design of singular lead compounds. Make-on-demand libraries have the advantage that candidates selected for follow-up experimental validation should be readily synthesizable, with some exceptions (10-20\% of compounds, anecdotally) due to imperfect enumeration rules. 

For drug discovery applications, virtual screening is often divided into two categories: structure-based \cite{Lyne2002,Ghosh2006,Cheng2012} and ligand-based {\cite{Willett2006,Ripphausen2011}}. Structure-based VS relies on scoring functions that relate information about a molecule and the target protein to binding affinity between the two. Docking analysis is a common paradigm in structure-based VS. Many software packages for this purpose exist, including AutoDock {\cite{S.1990}}, FlexX \cite{Rarey1996}, GOLD \cite{Jones1997}, and Glide \cite{Friesner2004}, and have been extensively reviewed  and compared \cite{Warren2006, Pagadala2017}. Ligand-based strategies, in contrast, make no direct consideration of the structure of the target protein. Instead, they rely on QSAR models and/or direct similarity assessments \cite{ballester_prospective_2009} that compare library compounds to a reference compound that exhibits desired properties. Many algorithms exist for making the similarity comparisons \cite{Cereto-Massague2015, Lo2018}.

Studies that validate virtual screening strategies by synthesizing and testing the compounds identified by their workflow include \cite{friedrich_from-complex_2016,fang_discovery_2016, lyu_ultra-large_2019}; we note that such validation is routine and expected in industrial drug discovery campaigns. In one example, \citeauthor{hoffer_integrated_2018} combine virtual screening with partially-automated synthesis and testing in a workflow for hit-to-lead optimization that they call diversity-oriented target-focused synthesis, or DOTS \cite{hoffer_integrated_2018, hoffer_covadots:_2019} (Figure~\ref{fig:hoffer_integrated_2018}). The DOTS framework begins with a hit fragment around which a virtual library is enumerated through the \emph{in silico} application of common synthetic reactions that combine the hit with commercially-available building blocks. Next, the authors apply their docking program, S4MPLE \cite{Hoffer2013}, which uses an evolutionary algorithm for conformational sampling, to select the compounds with favorable target interactions. Finally, the high-priority set is synthesized using a Chemspeed robotic synthesis platform to carry out expert-defined syntheses and subjected to {\it{in vitro}} evaluation. In one case study exploring optimization of an inhibitor of one of the two bromodomains of the BRD4 protein (which is implicated in inflammatory and cardiovascular diseases and cancer), all seventeen of the high-priority compounds had higher pIC50 values than the initial hit.

\begin{figure}[h]
  \centering
  \includegraphics[width=12cm]{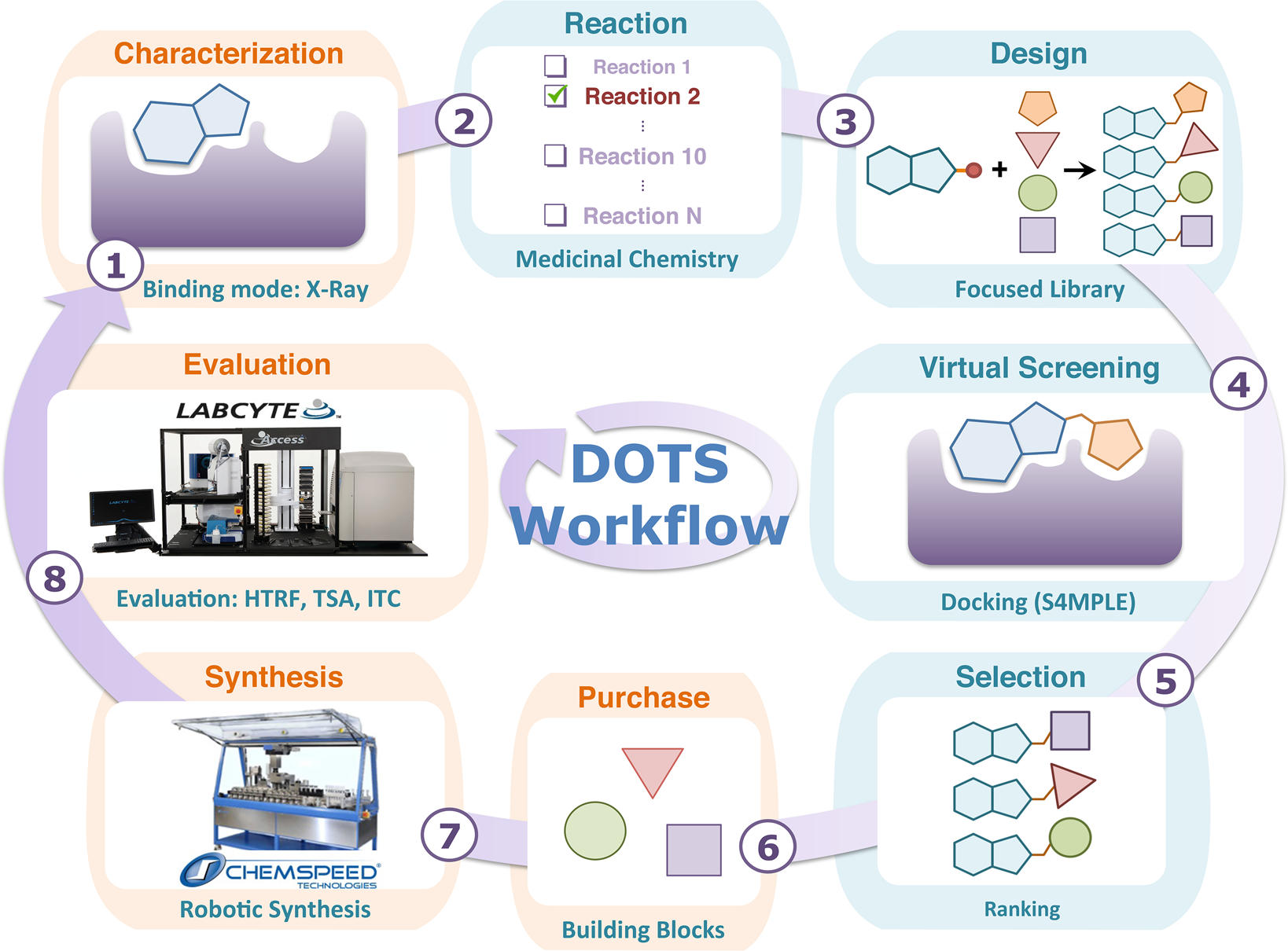}
  \caption{Workflow for diversity-oriented target-focused synthesis (DOTS). Experimental steps are in orange; computational steps are in blue. Figure reproduced from \citeauthor{hoffer_integrated_2018} \cite{hoffer_integrated_2018}.}
  \label{fig:hoffer_integrated_2018}
\end{figure}

\subsubsection{Discovery through molecular generation}
All of the case studies above examine search spaces defined by discrete sets of candidates. These candidate sets consist either of the compounds already existing in a database or library of interest, or they are somehow systematically enumerated. While some of these libraries are quite large, for example, the 170 million make-on-demand compounds from ref.~\citenum{lyu_ultra-large_2019} or the 11 billion in the REAL database \cite{enamine_REAL}, their discrete nature constrains the search space. Computational techniques such as deep generative models in which molecules are generated, manipulated, and/or optimized in a continuous latent space (Figure~\ref{fig:Gomez-Bombarelli2018}) have emerged and represent a means of overcoming the finiteness of discrete candidate sets (and, more specifically, as an alternative to earlier design approaches based, e.g., on genetic algorithms \cite{Schneider2005}). These models are predicated on the assumption that the generated compounds, by virtue of being drawn from the same distribution as the training molecules, will inherit the training molecules' important properties such as stability and synthesizability while being biased toward a specific property of interest (e.g., bioactivity) \cite{sanchez-lengeling_inverse_2018, elton_deep_2019}. Experimental validation is uniquely relevant for these techniques since they are not based on first-principles calculations, interpretable QSARs, or well-vetted heuristics, but rather neural models that create an obfuscated approximation of the distribution within chemical space and an underlying structure-function landscape.

\begin{figure}[h]
  \centering
  \includegraphics[width=12cm]{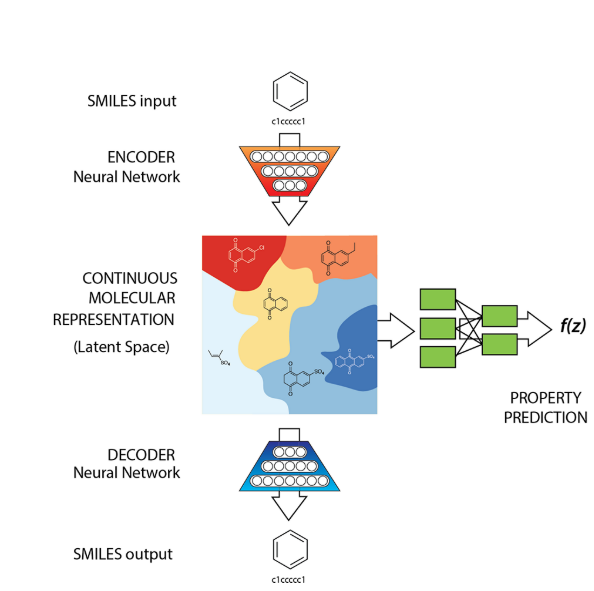}
  \caption{Diagram of an autoencoder for molecular discovery. Figure reproduced from \citeauthor{Gomez-Bombarelli2018} \cite{Gomez-Bombarelli2018}.}
  \label{fig:Gomez-Bombarelli2018}
\end{figure}

In an early example of the adaptation of deep generative networks to the pharmaceutical space, \citeauthor{Kadurin2017} describe the development of an adversarial autoencoder (AAE), which wraps an autoencoder in the generative adversarial network (GAN) training framework \cite{goodfellow2014generative} to identify antitumor agents based on  existing MCF-7 cell line assay data \cite{Kadurin2017} (see ref.~\citenum{Makhzani2015} for the original implementation and ref.~\citenum{kadurin_drugan_2017} for the improved technique, DruGAN). In another early example, \citeauthor{Gomez-Bombarelli2018} applied a VAE operating on SMILES strings (following the decoding approach of \citeauthor{bowman2015generating} \cite{bowman2015generating}) for the latent-space optimization of molecules with respect to druglikeness and synthetic accessibility metrics, demonstrating superior performance to random search and a genetic algorithm when initialized on low-performing molecules \cite{Gomez-Bombarelli2018}. 

Generative models with RNN encoding-decodings have emerged as one of the major paradigms in \emph{de novo} drug design \cite{Putin2018, Segler2018}. For example, \citeauthor{yuan_chemical_2017} use a character-level RNN \cite{karpathy2015multi} to generate virtual libraries of SMILES strings \cite{yuan_chemical_2017}. By training their model on 25,000 known VEGFR-2 inhibitors, they were able to generate a library enriched with high-affinity ligands relative to target-agnostic screening libraries, as judged through a computational docking program. Five of the highest-affinity ligands were selected for synthesis and testing and two were found to be more potent than vatalanib, a known inhibitor. \citeauthor{Bjerrum2017} adopted a similar approach using the ZINC12 database to train their model \cite{Bjerrum2017}. Their emphasis on evaluating the synthetic accessibility of the molecules that their model designed reflects the extent to which generative models have failed in this area historically. Combining this strategy with reinforcement learning to generate molecules that are similar to a seed compound \cite{Olivecrona2017}, molecules that have high predicted bioactivity against a particular target \cite{Olivecrona2017,Popova2018, Segler2018}, molecules that otherwise have desirable druglike properties (such as chemical beauty and Lipinski) \cite{Sanchez-Lengeling2017}, and molecules that represent an internally diverse set \cite{Putin2018} have proven fruitful, as have applications to peptide design \cite{muller_recurrent_2018, grisoni_designing_2018}. Transfer learning in the form of model pretraining has also been useful to successfully overcome the disadvantages inherent in low-data domains, for example to design modulators of therapeutically-relevant nuclear receptors \cite{Gupta2018,Merk2018}.

As \citeauthor{Jin2018} point out, a failing of the SMILES string representation in the molecular generation context is that a single molecule can usually be mapped to several distinct, valid SMILES strings, which complicates the creation of a latent space that varies smoothly from one molecule to another, similar one. They contribute an alternative approach, the \emph{junction tree variational autoencoder}, that generates molecular graphs rather than SMILES strings, demonstrating the ability to generate both a library of valid molecules as well as latent-space optimization to optimize molecules according to a joint logP-synthetic accessibility objective function \cite{Jin2018}.

Several of the generative model case studies cited herein include experimental validation {\cite{Merk2018,yuan_chemical_2017,Putin2018,putin_reinforced_2018,Polykovskiy2018, muller_recurrent_2018, grisoni_designing_2018}}, although the validation was not automated and the sets of generated molecules often required extensive filtering before  selection for synthesis. \citeauthor{Schneider2019} review fragment-based \emph{de novo} drug discovery efforts that specifically include experimental validation \cite{Schneider2019}. \citeauthor{Schneider2019} also highlight the fact that \emph{de novo} efforts are often plagued by synthesizability issues and advocate for the incorporation of CASP software into the workflow to help address this. In lieu of experimentation, some studies validate the capabilities of generative models by comparing distributions of properties of the generated molecules to those of the training set \cite{Ertl2017, Preuer2018}. See \cite{sanchez-lengeling_inverse_2018, schwalbe-koda_generative_2019, elton_deep_2019} for detailed reviews of the methods for and applications of generative models in chemistry and molecular design.

\subsection{\emph{Iterative} discovery of new physical matter}
\label{sec:cases:iterative_phys}
One rarely has a perfect understanding of a structure-function landscape, particularly in discovery applications where data can be limited. In this section, we focus on case studies in which  computer-assistance is applied to \emph{at least} the experimental selection aspect of an iterative discovery workflow (left half of Figure~\ref{fig:taxonomy_disc_phys_matt}).  In general, iterative discovery of new physical matter centers around a structure-function model that is used to reason about which experiments to perform. The results of the experiments are then used to devise the subsequent round and update the structure-function model such that more accurate predictions can be made. Iterative strategies like active learning and Bayesian optimization can be used to augment the set of available data efficiently, focusing on informative and/or promising experiments within the search space \cite{frazier2018tutorial, settles2012active}. Other iterative strategies include evolutionary algorithms, which operate by mutating candidates directly based on validation data from an experiment, a simulation, or a surrogate model. Model-free strategies where validation and feedback inherently drives experimental selection (e.g., directed evolution and continuous evolution techniques \cite{packer_methods_2015,scheuermann_dna-encoded_2006}) are out of the scope of this review, but do represent an important class of autonomous experimental platforms. 

\subsubsection{Discovery for pharmaceutical applications}

Given the complexity of the assays involved in characterizing new drug compounds, active learning is especially beneficial in the drug discovery domain to reduce the experimental burden and accelerate the search \cite{Murphy2011, reker_active-learning_2015}. Pool-based active learning (Figure \ref{fig:kangas_2014_poolbaseactivelearning}) refers to the selection of candidates from a discrete, pre-enumerated set of options \cite{settles2012active}. Active learning has been deployed in this setting both for information--to build accurate models of the activity of potential drug compounds against specified targets \cite{warmuth_active_2003,naik_efficient_2013}--as well as for performance--to identify active compounds in as few experiments as possible \cite{warmuth_active_2003,Fujiwara2008}. These aims are not mutually exclusive; in order to identify high-performing candidates, it is often the case that these algorithms must be designed to select experiments where activity is expected to be highest and experiments that support overall model accuracy, i.e., effectively balancing exploration and exploitation {\cite{kangas_efficient_2014,Reker2016,desai_rapid_2013, li_designing_2019}}. 

\begin{figure}[h]
  \centering
  \includegraphics[width=12cm]{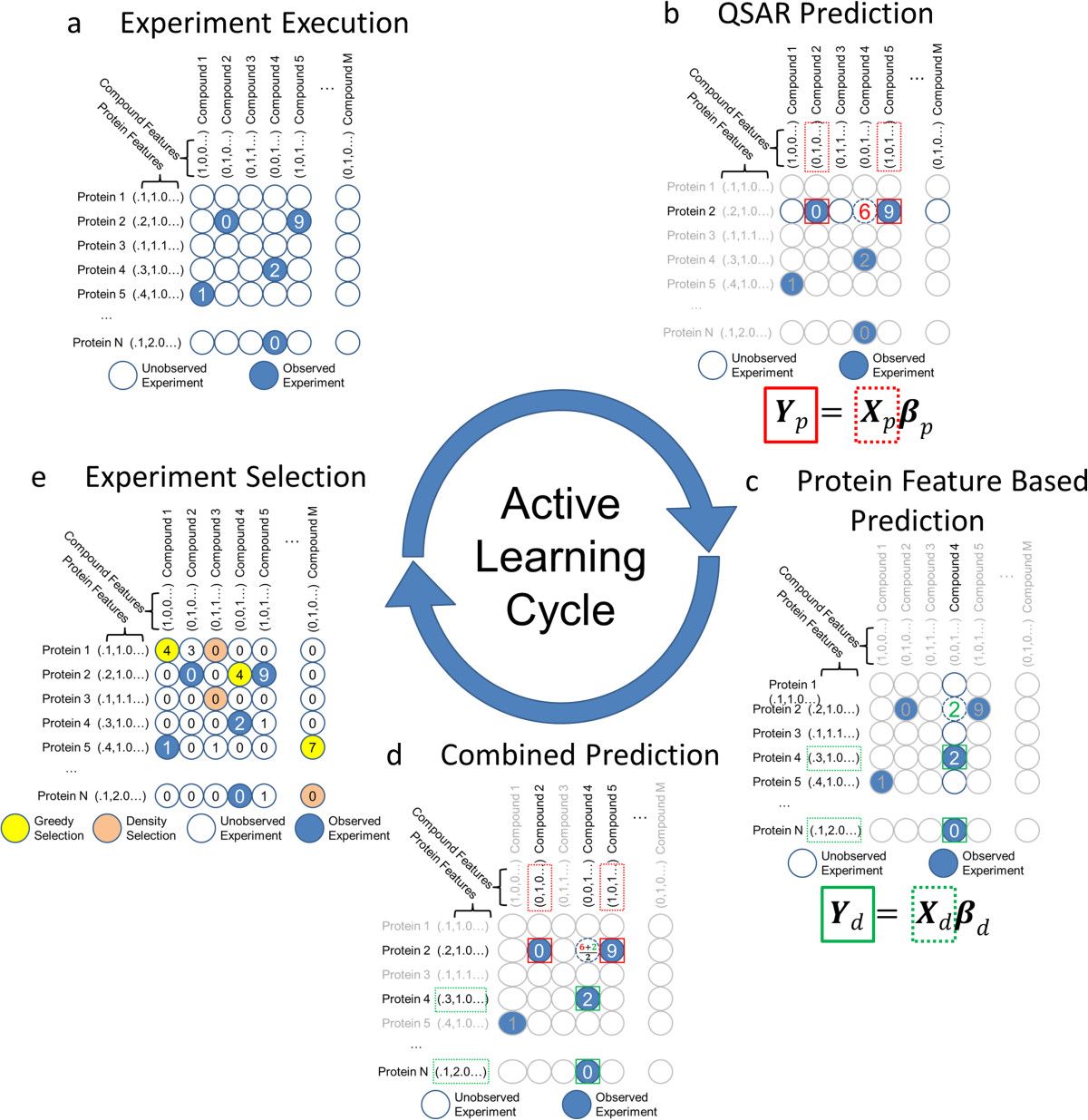}
  \caption{Workflow for pool-based active learning to identify compounds that bind strongly to proteins within an $N$ compound $\times$ $M$ protein space of interactions. Figure reproduced from \citeauthor{kangas_efficient_2014} \cite{kangas_efficient_2014}.}
  \label{fig:kangas_2014_poolbaseactivelearning}
\end{figure}

A particularly noteworthy example of the pool-based active learning strategy is the platform Eve, which was designed to conduct autonomous, closed-loop hit identification \cite{bilsland_yeast-based_2013,williams_cheaper_2015}. Experimentally, Eve is endowed with the capacity to rapidly screen predefined compound libraries against a variety of biological assays at a rate of >10,000 compounds per day. The platform uses collected data to create a surrogate model of the structure-activity landscape and then selects subsequent rounds of compounds with high predicted activity, selectivity, and/or prediction variance, rather than exhaustively exploring its search space. The authors created an econometric model of the drug discovery process that accounts for (a) the utility of a hit (b) the utility of the reduction in experimental space that needs to be screened, and (c) the cost of missed hits, among others, and found that it is typically more economical to use active learning than  brute-force screening. The success of Eve depends on access to a large library of compounds that contains at least one acceptably-high-performing compound.

Increasing the size of the search space increases the likelihood there is a high-performing global optimum (although it may be difficult to identify). In experimental settings, the ability to synthesize compounds on-demand expands the search space beyond the set of in-stock compounds. \citeauthor{desai_rapid_2013} describe a microfluidic platform able to produce $27 \times 10$ compounds on-demand via a one-step Sonogashira coupling \cite{desai_rapid_2013}. Impressively, the platform integrates synthesis with online purification, dilution, and activity assay against Abl1 and Abl2 kinases. A random forest model was created to approximate the structure-activity landscape and guide experiment selection. Experiments were chosen via one of two approaches--one greedy approach to maximize expected activity and one to explore undersampled chemical space--and results were used to update the surrogate model  \cite{czechtizky_integrated_2013}. Despite the expansion in the search space achieved by incorporating synthesis capabilities, the design space explored in this example remains extremely narrow; a brute-force search would have been tractable and potentially faster if parallelized. Still, this study is an excellent proof-of-concept for closed-loop synthesis, purification, and testing. More flexible  synthesis-purification and synthesis-purification-testing platforms have been developed but only applied to open-loop discovery with manual compound design and synthesis planning \cite{godfrey_remote-controlled_2013, hochlowski_integrated_2011, baranczak_integrated_2017, gesmundo_nanoscale_2018} (Figure~\ref{fig:baranczak_integrated_2017}). Table 2 in ref.~\citenum{chow_streamlining_2018} reviews some additional examples of integrated synthesis and testing.

\begin{figure}[h]
  \centering
  \includegraphics[width=\twocolumnsize]{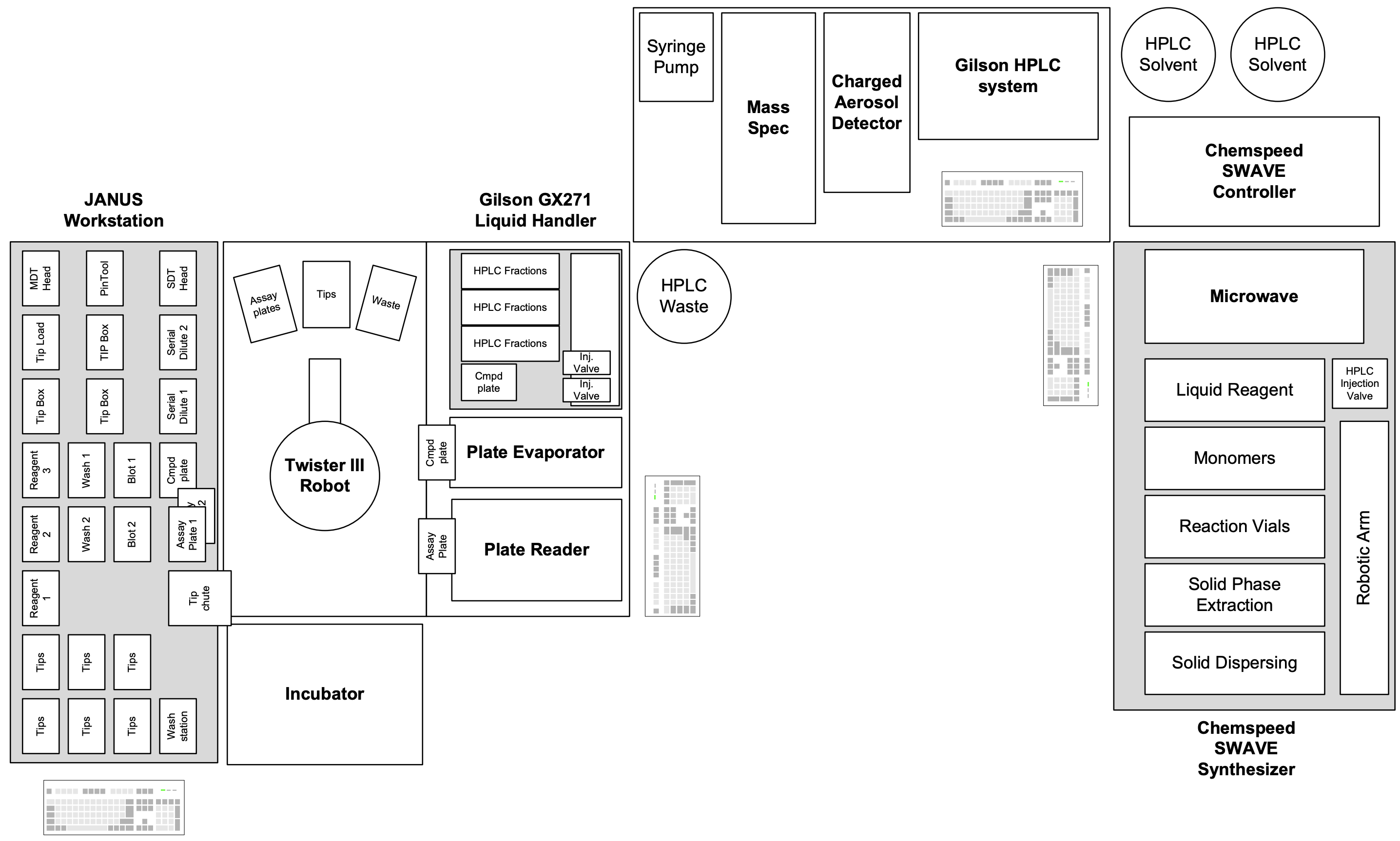}
  \caption{Integrated platform for open-loop synthesis, purification, and testing from AbbVie. Figure reproduced from \citeauthor{baranczak_integrated_2017} \cite{baranczak_integrated_2017}.}
  \label{fig:baranczak_integrated_2017}
\end{figure}

Evolutionary strategies are another means of expanding the search space beyond a set of pre-enumerated candidates. \citeauthor{besnard_automated_2012} employ one such technique in which candidate compounds evolve as part of the iterative process \cite{besnard_automated_2012}. Specifically, on each iteration, new candidates are evolved from the highest performers from the previous generation through the application of structural transformations from the medicinal chemistry literature. Here, the discovery workflow relies on the development of accurate QSAR models trained on ChEMBL data to ensure alignment between the \emph{in silico} performance and experimental activity. \citeauthor{firth_moarf_2015} use a related technique that evolves molecules using a fragment replacement routine (RATS, rapid alignment of topological scaffolds), which enables a less constrained exploration of chemical space than does reaction enumeration; this approach was demonstrated on a surrogate multi-objective function (including a model of CDK2 activity) starting from a known active scaffold, with manual experimental validation of a small library of recommendations \cite{firth_moarf_2015}.

Genetic algorithms (GAs) are related to the approaches used by \citeauthor{besnard_automated_2012} and \citeauthor{firth_moarf_2015}. Candidate compounds (experiments) are proposed as mutations of a parent compound whose performance is known; allowed mutations serve as a constraint on the search space and the optimization trajectory \cite{venkatasubramanian_computer-aided_1994} (see Figure~\ref{fig:Dey2008} for an illustration of one implementation and the corresponding algorithm flowchart). In contrast to active learning, GAs tend to use \emph{static} fitness functions for compound scoring (although a few rely on experimental outcomes instead \cite{pickett_automated_2011}). In a very early example of iterative molecular optimization using a genetic algorithm, \citeauthor{weber_optimization_1995} describe the identification of inhibitors of the serine protease thrombin; 16 generations led to the identification of sub-micromolar inhibitors \cite{weber_discovery_1999}. The key to this approach is that the $10 \times 40 \times 10 \times 40$ design space was defined by discrete substrate choices in a 4-component Ugi-type reaction to ensure straightforward (albeit manual) synthesis and testing. Other iterative strategies for drug discovery include \emph{in silico} application of synthetic transformations to generate molecules that are scored based on their similarity to a target molecule and are, in principle, synthetically-accessible \cite{Button2019}.

\begin{figure}[h]
  \centering
  \includegraphics[width=12cm]{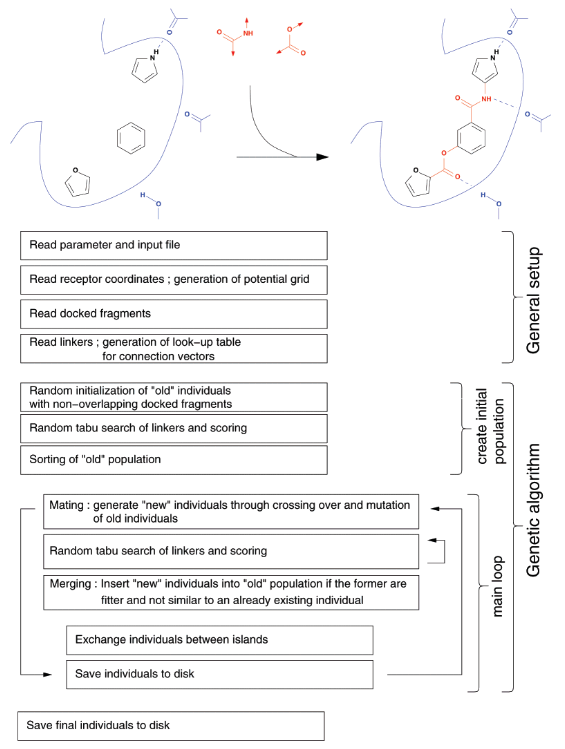}
  \caption{(Top) Schematic illustration of one approach to genetic algorithm-based drug design: predocked fragments (black) are linked to fragments (red) from a user-supplied list, with the target protein and its polar groups indicated in blue. (Bottom) The flow chart of the algorithm. Figure reproduced from \citeauthor{Dey2008} \cite{Dey2008}.}
  \label{fig:Dey2008}
\end{figure}

GAs have been successfully used to conduct both single- and multi-objective optimizations that account for factors including target protein binding affinity \cite{Wang2000,Pegg2001}, cost and bioavailability \cite{Gillet2002}, and similarity to a chosen compound {\cite{Sheridan1995,Gillet2002,Schneider2000,Dey2008}}. Many strategies are fragment-based, operating in the manner described above, although some also allow atom-level mutations {\cite{Douguet2000}}. Strategies that operate directly on molecular graphs have also been proposed {\cite{Brown2004}}, and in the case of peptide design, one can operate directly on sequences {\cite{Kamphausen2002}}. Despite the increased interest in deep learning techniques, genetic algorithms remain a powerful strategy for exploring chemical space \cite{jensen_graph-based_2018}. A number of reviews describe drug discovery applications of GAs and evolutionary algorithms more generally \cite{Nicolaou2013, Clark1996, Schneider2005, Terfloth2001, Nicolaou2007}.

\subsubsection{Discovery for materials applications}
Iterative experimental design strategies are increasingly being adopted to guide discovery in the materials space. For example, \citeauthor{Xue2016} use the Bayesian optimization framework EGO \cite{Jones1998} to select experiments from a discretized composition space that ultimately led them to NiTi-based shape memory alloys delivering low thermal hysteresis \cite{Xue2016}. Similar approaches have been used for the same \cite{solomou_multi_2018} and other applications to discover \ce{BaTiO3}-based piezoelectric materials with large electrostrains \cite{Yuan2018} and to optimize melting temperature \cite{Seko2014}. Compared to other domains, materials discovery tends to be relatively conducive to fully computational approaches since the properties we can calculate are more directly relevant to the functions we wish to optimize; as a result, several studies have used Bayesian optimization to select compounds for evaluation with calculations or simulations, rather than experiments, e.g., to optimize thermal conductivities \cite{Seko2015} and elastic moduli \cite{Balachandran2016}.  Additionally, some instances of generative models--described in the noniterative discovery section above--incorporate Bayesian optimization to optimize compounds for desirable performance \cite{Gomez-Bombarelli2018}.  

Related active learning strategies have been deployed in materials development as well. For example, \citeauthor{tran_active_2018} use DFT to validate candidates proposed through active learning from a fixed pool of 1,499 intermetallic compounds from the Materials Project, with the goal of identifying high-performance electrocatalysts for \ce{CO2} reduction and \ce{H2} evolution \cite{tran_active_2018}. \citeauthor{gubaev2019accelerating} use active learning to efficiently create a DFT surrogate for predicting the convex hulls of metallic alloy systems, ultimately discovering previously unknown stable alloys \cite{gubaev2019accelerating}. Even iterative greedy searches have proven effective in prioritizing simulations of a fixed library of candidate materials for hydrogen storage \cite{thornton_materials_2017}.

Several recent studies combine GAs, surrogate models for electronic structure calculations, and active learning for the discovery of spin crossover complexes and transition metal catalysts \cite{janet_accelerating_2018, nandy_strategies_2018} (Figure~\ref{fig:nandy_strategies_2018}). Treating the calculations as reliable measures of performance, these studies represent fully autonomous discovery within the space of organometallic complexes the genetic algorithm is able to explore.  Among the many additional applications of GAs to materials discovery have been polymer design \cite{Venkatasubramanian1994}, the identification of stable four-component alloys \cite{johannesson_combined_2002}, promising polymers for OPVs \cite{oboyle_computational_2011}, MOFs for carbon capture \cite{Chung2016}, and polymer dielectrics with user-defined dielectric constants and bandgaps \cite{Mannodi-Kanakkithodi2016}.  An excellent review of applications of active learning and Bayesian optimization to materials development is provided in ref.~{\citenum{Lookman2019}}.  

\begin{figure}[h]
  \centering
  \includegraphics[width=6cm]{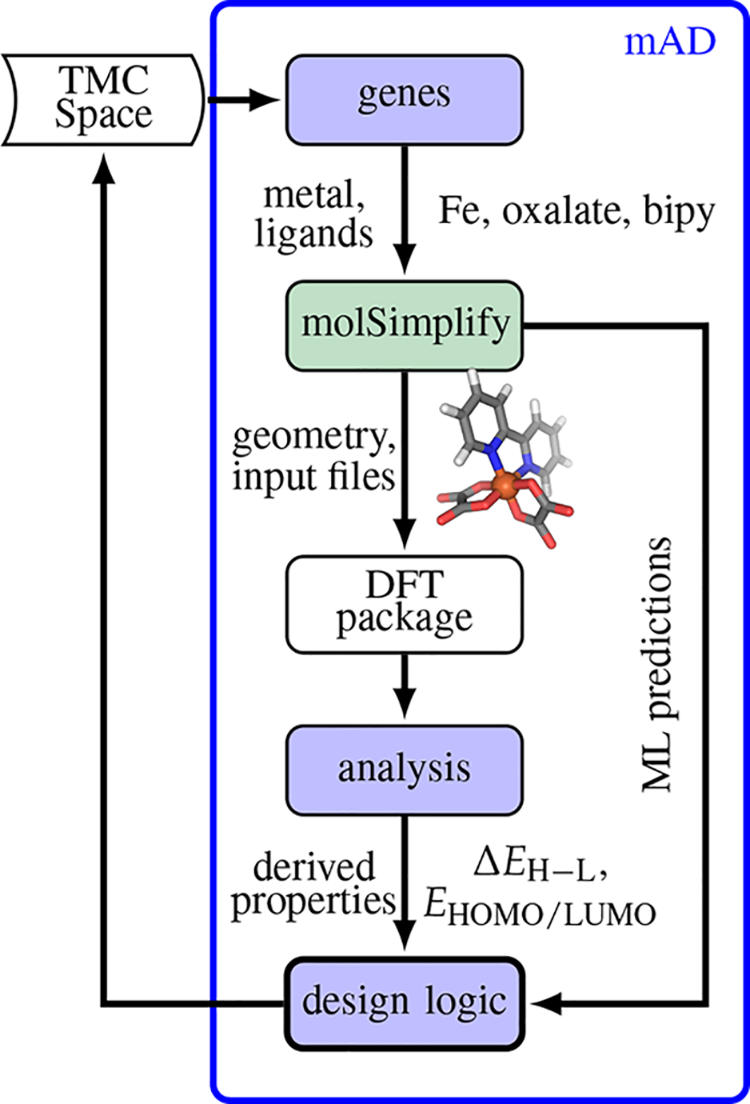}
  \caption{Workflow for the iterative design of transition metal complexes (TMCs) using a genetic algorithm and automated DFT calculations. Figure reproduced from \citeauthor{nandy_strategies_2018} \cite{nandy_strategies_2018}.}
  \label{fig:nandy_strategies_2018}
\end{figure}

The utility of computational validation for materials discovery is somewhat offset by the complexity of experimental validation. As stated above, the synthesis and analysis of materials and devices can be difficult to automate. A recent study by \citeauthor{macleod_self-driving_2019} (Figure~\ref{fig:macleod_2019_ada}) serves as an excellent example of automating more than simple solution-phase chemistry \cite{macleod_self-driving_2019}. Precursor solutions are spin cast into thin films, which are thermally annealed and analyzed for their optical and electronic properties. The platform, Ada, uses ChemOS \cite{roch_chemos:_2018} for hardware orchestration and Pheonics \cite{hase_phoenics:_2018} for Bayesian optimization to explore a two dimensional continuous search space of composition and annealing temperature; its objective is to optimize a pseudomobility metric that correlates with hole mobility. While what Ada measures is still a proxy for the true performance of a multilayer device, the ability to miniaturize and automate fabrication processes like thin film casting expands the scope of problems able to be tackled by autonomous platforms.

\begin{figure}[h]
  \centering
  \includegraphics[width=13cm]{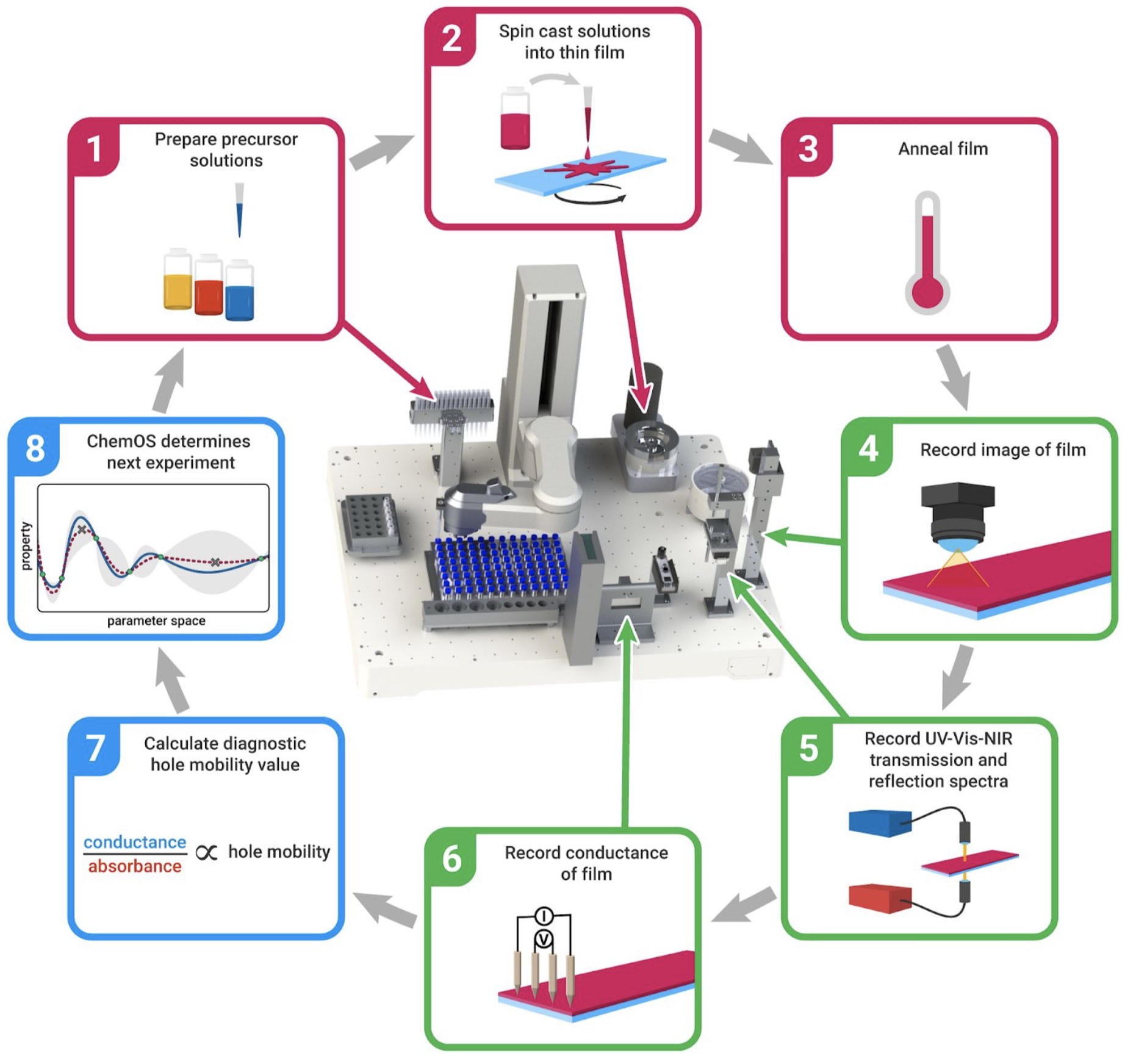}
  \caption{The autonomous platform Ada for optimizing optoelectronic properties of spin cast thin films. Figure reproduced from \citeauthor{macleod_self-driving_2019} \cite{macleod_self-driving_2019}.}
  \label{fig:macleod_2019_ada}
\end{figure}

\FloatBarrier

\subsection{Brief summary of discovery in other domains}
\label{sec:cases:brief_summary}

There are many more attempts to automate aspects of the discovery process and incorporate machine learning into scientific workflows than can be mentioned here. Some pertinent to biology and human health are summarized below. A more comprehensive, collaborative review can be found in ref.~\citenum{ching_travers_opportunities_2018}.

\textbf{To identify correlations from text mining: } 
The enormous size of the scientific literature makes it challenging to analyze holistically. Information retrieval tools are needed to bring together relevant pieces of information, either for automated analysis or manual inspection. ARROWSMITH was an early system for the latter use \cite{swanson_interactive_1997} that identified MEDLINE abstracts with overlapping terms potentially indicating an implicit causal relationship. This was used to discover a number of testable hypotheses including a link between magnesium and migraines \cite{swanson_migraine_1988}. Literature mining is an essential tool for organizing biological data and enabling computational studies \cite{muller_textpresso:_2004, krallinger_text-mining_2005,jensen_literature_2006, krallinger_linking_2008, gyori_word_2017, sayers_database_2019}.

\textbf{To identify trends in genomics data: }
The vast quantity of structured genetic information brought about by the Human Genome Project and advances in DNA sequencing is well-suited for data mining. As one example, probabilistic graph models can be built from genetic, protein-interaction, and metabolic pathway information to propose hypotheses for gene functions \cite{leach_biomedical_2009,Bell2011}. Two practical introductions to machine learning for genomics can be found in ref.~\citenum{libbrecht_machine_2015} and ref.~\citenum{eraslan_deep_2019}.

\textbf{To engineer new proteins: }
Protein engineering through directed evolution requires navigating a high dimensional and discontinuous structure-function landscape. Random mutagenesis navigates this space blindly, one step at a time, while site-directed mutagenesis requires knowledge of which amino acid positions to perturb. Supervised machine learning models and other statistical techniques can assist in the selection of mutants over multiple rounds of evolution, particularly when dealing with nonadditive effects, as reviewed by \citeauthor{yang_machine_2018} \cite{yang_machine_2018}. Computational techniques are used for protein engineering in many other ways \cite{verma_computer-aided_2012} including protein structure prediction \cite{evans_novo_2018, alquraishi_end--end_2019}.

\textbf{To identify gene/enzyme relationships: }
A high-profile example of an automated platform for molecular genetics is \citeauthor{king_functional_2004}'s Adam \cite{king_functional_2004,king_automation_2009,king_automating_2018}. In the original demonstration, Adam was made aware of the aromatic amino acid synthesis pathway in yeast, hypothesized which of 15 open reading frames (ORFs) encoded which enzyme, and selected growth experiments to perform (choosing one knockout mutant out of 15 options and one to two metabolites out of 9 options). While this almost represented a truly closed-loop system, there were still manual steps involved in transferring well plates between the liquid handler, incubator, and plate reader. Additionally, these experimental and hypothesis spaces are extraordinarily narrow: the model's accuracy using an active search strategy was 80.1\% compared to 74.0\% for a naive method that chose the cheapest experiment yet to be performed.  In ref.~\citenum{king_automating_2018}, King acknowledges the criticism that ``the new scientific knowledge was implicit in the formulation of the problem, and is therefore not novel''.

\fi

\section{Conclusion}
In the first part of this review, we have defined three broad categories of discovery--physical matter, processes, and models--and suggested guidelines for evaluating the extent to which a scientific workflow can be described as autonomous: (i) How broadly is the goal defined? (ii) How constrained is the search/design space? (iii) How are experiments for validation/feedback selected? (iv) How superior to a brute force search is navigation of the design space? (v) How are experiments for validation/feedback performed? (vi) How are results organized and interpreted? (vii) Does the discovery outcome contribute to broader scientific knowledge?

As illustrated by the case studies we have included, there has been substantial progress in developing methods that build toward autonomous discovery. Yet there are few examples of true closed-loop discovery for all but the narrowest design spaces--often a consequence of the complexity of automating experimental validation.

We continue to face both practical and methodological challenges in our quest for autonomous discovery. The second part of this review will reflect on a selection of case studies in terms of the questions we pose and then describe remaining challenges where further development is required.

\section{Acknowledgements}
We thank Thomas Struble for providing comments on the manuscript and our other colleagues and collaborators for useful conversations around this topic. This work was supported by the Machine Learning for Pharmaceutical Discovery and Synthesis Consortium and the DARPA Make-It program under contract ARO W911NF-16-2-0023.

\iffull 
\singlespacing
\printbibliography
\fi

\end{document}